\documentclass[a4paper]{article}
\usepackage[top=2.5cm,left=2.5cm,right=2.5cm,bottom=1.5cm]{geometry}
\usepackage{authblk}             
\usepackage{cite}
\usepackage[obeyspaces]{url}     
\usepackage[parfill]{parskip}    
\usepackage{color}
\usepackage{graphicx}
\usepackage{amsmath}
\usepackage{amsfonts}
\usepackage[utf8]{inputenc}      
\usepackage{mathtools}  
\usepackage{calrsfs}             
\usepackage{xfrac}               
\usepackage{braket}              

\usepackage{wrapfig}
\usepackage{commath}
\usepackage{bm}
\usepackage{units}
\usepackage[usenames,dvipsnames,svgnames,table]{xcolor}

\usepackage[percent]{overpic}

\DeclareMathOperator{\arctanh}{arctanh}
\DeclareMathOperator{\sech}{sech}
\DeclareMathOperator{\gd}{gd}

\renewcommand{\Re}{\operatorname{Re}}

\newcommand{\pa}{\partial_1}
\newcommand{\pb}{\partial_2}

\newcommand{\pt}{\partial_t}

\newcommand{\prv}{\nabla_{\vec{r}}}

\newcommand{\vp}{\varphi}

\newcommand{\matrixel}[3]{\left< #1 \vphantom{#2#3} \right| #2 \left| #3 \vphantom{#1#2} \right>}

\renewcommand{\vec}{\bm}

\title{The Geometric Potential of the Exact Electron Factorization: Meaning, significance and application}
\author[1]{Jakub Koc\'ak}
\author[2]{Eli Kraisler}
\author[1]{Axel Schild}
\affil[1]{Laboratorium für Physikalische Chemie, ETH Z\"urich, Vladimir-Prelog-Weg 2, 8093 Z\"urich, Switzerland}
\affil[2]{Fritz Haber Center for Molecular Dynamics, Institute of Chemistry, The Hebrew University of Jerusalem, 91904 Jerusalem, Israel}
\date{}

\begin{document}
  
  \maketitle
  
  \begin{abstract}
    The theoretical and computational description of materials properties is a task of utmost scientific and technological importance. 
    A first-principles description of electron-electron interactions poses an immense challenge that is usually approached by converting the many-electron problem to an effective one-electron problem. 
    There are different ways to obtain an exact one-electron theory for a many-electron system.
    An emergent method is the exact electron factorization (EEF) -- one of the branches of the Exact Factorization approach to many-body systems. In the EEF, the Schr\"odinger equation for one electron, in the environment of all other electrons, is formulated. The influence of the environment is reflected in the potential $v^{\rm H}$, which represents the energy of the environment, and in a potential $v^{\rm G}$, which has a geometrical meaning. 
    In this paper, we focus on $v^{\rm G}$ and study its properties in detail. 
    We investigate the geometric origin of $v^{\rm G}$ as a metric measuring the change of the environment, exemplify how translation and scaling of the state of the environment are reflected in $v^{\rm G}$, and explain its shape for homo- and heteronuclear diatomic model systems.
    Based on the close connection between the EEF and density functional theory, we also use $v^{\rm G}$ to provide an alternative interpretation to the Pauli potential in orbital-free density functional theory.
  \end{abstract}
  
  \section{Introduction}
  \label{sec:intro}
  
  The quantum-mechanical solution of the many-electron problem is difficult but necessary to determine the properties of molecules and materials, as well as to predict the outcomes of chemical reactions \cite{dirac1929,motta2017}.
  A common strategy to solve the many-electron problem is to turn it into a one-electron problem, and the arguably most famous and most successful representative of this strategy is density functional theory (DFT) \cite{hohenberg1964,dreizler1990,ullrich2013}, in particular Kohn-Sham (KS) DFT \cite{kohn1965}.
  The central idea of KS-DFT is to map an interacting many-electron system to a fictitious system of non-interacting electrons, the KS system, such that both systems have the same one-electron density $\rho(\vec{r})$.
  As the electrons in the KS system are non-interacting, the many-electron problem is effectively reduced to a one-electron problem.
  To determine the KS system, the one-electron KS potential $v^{\rm KS}(\vec{r})$ is needed, which is typically treated as a functional of $\rho$ or of the KS orbitals, i.e., of the eigenfunctions of $v^{\rm KS}$.
  The functional dependence is not completely known, but suitable approximations allow to answer many questions of physical and chemical relevance \cite{mohr2015}.
  
  However, there are potentially useful alternatives to KS-DFT.
  One such alternative is the Exact Electron Factorization (EEF) \cite{schild2017,kocak2020,complet}.
  The EEF is based on the idea that a multi-particle wavefunction can be separated into a product of a marginal and a conditional part. 
  This idea was first applied to the molecular wavefunction to separate the nuclear wavefunction from the electronic wavefunction \cite{hunter1975} in the spirit of the Born-Oppenheimer approximation \cite{born1927}, but without actually making an approximation.
  Recently, the formalism of the wavefunction separation has been further developed as exact factorization \cite{abedi2010,abedi2012,gonze2018} and has, for example, become the basis of a mixed quantum-classical algorithm to compute molecular quantum dynamics \cite{curchod2018,agostini2019}.
  
  In the context of the many-electron problem, the same idea is used in the EEF to reduce a system of $N$ electrons to a one-electron problem:
  The interacting $N$-electron wavefunction $\psi(\vec{r}_1,\dots,\vec{r}_N)$ is written as a product of a marginal one-electron wavefunction $\chi(\vec{r}_1)$ and a conditional wavefunction $\phi(\vec{r}_2,\dots,\vec{r}_N;\vec{r}_1)$ \cite{hunter1986}.
  Then $\chi$ corresponds to the exact one-electron density,  $|\chi(\vec{r}_1)|^2 = \rho(\vec{r}_1)$, and is an eigenstate of a one-electron Schr\"odinger equation.
  The conditional wavefunction $\phi(\vec{r}_2,\dots,\vec{r}_N;\vec{r}_1)$ depends on the coordinates of $N-1$ electrons, and, also, parametrically on the coordinates of the remaining electron of the $N$-electron system.
  
  The EEF provides a simple physical picture of the resulting one-electron description of the many-electron problem as that of one electron (with wavefunction $\chi$) in the environment of the other electrons (with wavefunction $\phi$). 
  The effective one-electron potential $v$ that appears in the one-electron Schr\"odinger equation determining $\chi$ is given by
  \begin{align}
    v(\vec{r}) = v^{\rm H}(\vec{r}) + v^{\rm G}(\vec{r}),
    \label{eq:veef}
  \end{align}
  where $v^{\rm H}$ is the energy of the environment and $v^{\rm G}$ is the so-called geometric potential.
  Also, from the EEF formalism it is immediately clear that there is a gauge freedom in the theory and that a vector potential $\vec{A}(\vec{r})$ can also appear in addition to the scalar potential $v(\vec{r})$, as detailed below.
  
  The EEF is closely connected to DFT.
  In particular, it was well-known that the amplitude $\sqrt{\rho(\vec{r})} = |\chi(\vec{r})|$ of the one-electron density is an eigenstate of a one-electron Schr\"odinger equation with the effective potential $v(\vec{r})$  
  \cite{levy1984,march1986,march1987,levy1988,kraisler2020rev}.
  This fact is the basis of orbital-free (OF) DFT, a method that aims at directly approximating $v(\vec{r})$ and that promises to be computationally very efficient for large systems \cite{wang2002,karasiev2012}.
  However, $v(\vec{r})$ is often analyzed in terms of the KS system, and the effective potential is written as 
  \begin{align}
    v(\vec{r}) = v^{\rm P}(\vec{r}) + v^{\rm KS}(\vec{r}), 
    \label{eq:pauli}
  \end{align}
  where $v^{\rm P}$ is the so-called Pauli potential. 
  This leads to two challenges: Approximating $v^{\rm XC}$, which is the main problem of KS-DFT, and approximating $v^{\rm P}$. 
  There has still been limited success in approximating $ v^{\rm P}$, but recently its properties received increased attention \cite{deb1983,perdew2007,karasiev2009,karasiev2013,karasiev2015,finzel2016,finzel2017,constantin2018,finzel2018a,finzel2018b,finzel2019,smiga2020,kraisler2020}.
  
  The connection between $\phi$ and the potentials $v$, $v^{\rm KS}$, and $v^{\rm P}$ has been of interest for a while and studies exits for atoms \cite{buijse1989,gritsenko1994,leeuwen1994,leeuwen1995} and diatomics \cite{buijse1989,gritsenko1996a,gritsenko1996b,gritsenko1997}, as well as further appearances of $\phi$ in the DFT literature \cite{helbig2009,saavedra2016,giorgi2016b,giarrusso2018,kumar2019,kumar2020,giarrusso2020,mccarty2020}.
  In a recent article \cite{complet}, we used the connection between $\phi$ and $v$ as revealed by the EEF formalism to explain the appearance of steps in $v$.
  These steps are charge-transfer steps that originate from a rearrangement of the electrons in the environment such that the energy of the environment $v^{\rm H}$ changes.
  Due to the connection of the EEF with KS-DFT, is also became clear that the mechanism underlying similar charge-transfer steps in $v^{\rm KS}$ \cite{hodgson2017} is the same.
  
  While the meaning of $v^{\rm H}$ is straightforward, the meaning of the geometric potential $v^{\rm G}$ in the EEF is less obvious.
  In this article, we study $v^{\rm G}$ and shed light on its interpretation and its properies.
  For this purpose, in Sec.\ \ref{sec:eef} we first present the theory of the EEF and its connection to OF-DFT.
  The EEF is based on the many-electron wavefunction and replacing it with the KS wavefunction provides the connection between the two decomposition of $v$ in \eqref{eq:veef} and in \eqref{eq:pauli}.
  From this, we learn that the geometric potential $v^{\rm G}$ is typically a part of the Pauli potential $v^{\rm P}$, and we obtain an interpretation of $v^{\rm P}$ that is different from the standard one.
  Thereafter, in Sec.\ \ref{sec:geopot} we explain why $v^{\rm G}$ is called the geometric potential.
  We show how $v^{\rm G}$ and the vector potential $\vec{A}$ that also appears in the EEF are related to the quantum geometric tensor and how this tensor encodes the geometric properties of the environment.
  Also, we study how simple changes of the shape of the wavefunction of the environment are related to the properties of $v^{\rm G}$, and we analyze the geometric aspects using a two-state model.
  Third, in Sec.\ \ref{sec:dia}, we study how $v^{\rm G}$ behaves for a numerically solvable model of a two-electron homo- and heteronuclear diatomic molecule in one dimension.
  With the help of the two-state model, a quantitative analysis of $v^{\rm G}$ becomes possible and we can show explicitly how changes of the environment are encoded in $v^{\rm G}$ for that model.
  Our article then closes in Sec.\ \ref{sec:sum} with a short summary of the findings and ideas for future research.
  
  \section{The exact electron factorization}
  \label{sec:eef}
  
  In this section, the EEF \cite{schild2017,kocak2020,complet} is presented as an exact way to reduce an $N$-electron problem to a one-electron problem.
  The resulting one-electron problem is that of one electron being subject to a scalar potential $v(\vec{r})$ and a vector potential $\vec{A}(\vec{r})$, together representing the environment, i.e. the other electrons.
  The solution of the EEF one-electron Schr\"odinger equation with these potentials yields the exact one-electron density $\rho(\vec{r})$ and the corresponding current density $\vec{j}(\vec{r})$, as well as one-electron observables of the many-electron system.
  
  \subsection{Formalism of the EEF}
  
  In the following, we consider a system of $N$ non-relativistic spinless interacting electrons in the Born-Oppenheimer approximation \cite{born1927}.
  We use Hartree atomic units and consider the ground state of the system, i.e., the energetically lowest fully antisymmetric eigenstate $\psi(\vec{r}_1,\dots,\vec{r}_N)$ of the many-electron Hamiltonian for some external potential $v^{\rm ext}$.
  The generalization to include electron spin and excited states is straightforward but complicates the presentation, hence it is not discussed here.
  For brevity, we sometimes substitute the electronic coordinates with numbers, e.g.\ $\psi(\vec{r}_1,\dots,\vec{r}_N) \equiv \psi(1,\dots,N)$.
  
  The many-electron Schr\"odinger equation is
  \begin{align}
    \left( -\sum_{j=1}^N \frac{\nabla_j^2}{2} + V(1,\dots,N) \right) \psi(1,\dots,N) = E \psi(1,\dots,N)
    \label{eq:se}
  \end{align}
  with the scalar potential 
  \begin{align}
    V(1,\dots,N) = \sum_{j=1}^N v^{\rm ext}(j) + \sum_{j=1}^N \sum_{k=j+1}^N v_{\rm ee}(j,k)
    \label{eq:vel}
  \end{align}
  that is the sum of one-electron (external) potentials $v^{\rm ext}(\vec{r})$, which include the attraction of the electrons to the nuclei, and the electron-electron interaction $v_{\rm ee}(\vec{r}_j,\vec{r}_k)$. For the exact case this is Coulomb repulsion $v_{\rm ee}(\vec{r}_j,\vec{r}_k) = 1/\left|\vec{r}_j -\vec{r}_k \right|$, but below we also consider other model potentials for $v_{\rm ee}$.
  
  The EEF \cite{schild2017} is based on the basic fact from the theory of probability that a joint probability can be written as product of a marginal and a conditional probability \cite{hunter1975,abedi2010}.
  In terms of wavefunctions, this translates to 
  \begin{align}
    \psi(1,\dots,N) = \chi(1) \phi(2,\dots,N;1).
    \label{eq:eef_ansatz}
  \end{align}
  $\chi(1)$ is the probability amplitude of finding any electron at $\vec{r}_1$, whereas  $\phi(2,\dots,N;1)$ is the conditional probability amplitude of finding $N-1$ electrons at $\vec{r}_2, \dots, \vec{r}_N$, given there is an electron at $\vec{r}_1$.
  
\begin{figure*}[!htbp]
    \centering
    \includegraphics[width=.9\linewidth]{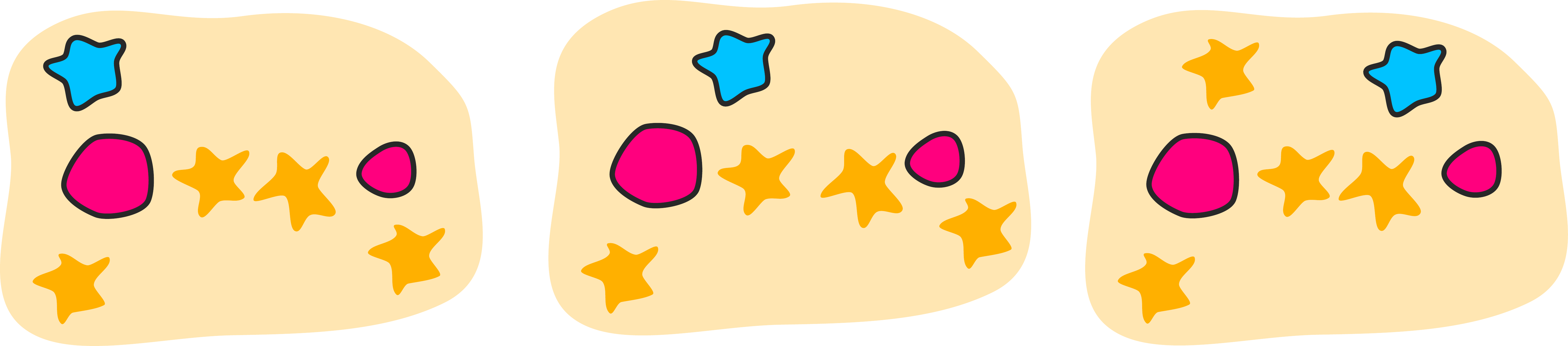}
    \caption{
    Idea of the exact electron factorization illustrated with a cartoon representing a diatomic molecule.
    The two round (magenta) shapes represent two nuclei, the five star-like shapes (yellow and blue) represent electrons.
    If the shape has a black border (the nuclei and the blue electron), its position is a condition:
    The nuclei are clamped and the state of the four yellow electrons (the environment) is described by the conditional wavefunction $\phi(1,2,3,4;\vec{r})$ for a given position $\vec{r}$ of the blue electron.
    The blue electron feels the external potential $v^{\rm ext}(\vec{r})$ that describes the interaction with the nuclei, the potential $v^{\rm H}(\vec{r})$ that is the energy of the 4-electron system given another electron is at $\vec{r}$, and the potential $v^{\rm G}(\vec{r})$ that can be thought of as the additional energy needed to change the state of the 4-electron system when the position $\vec{r}$ of the blue electron is changed.    
    A possible vector potential $\vec{A}(\vec{r})$ might also be felt by the blue electron, e.g., if the system is rotating.
    We emphasize that the blue electron is any electron and the fermionic antisymmetry conditions are unbroken.
    }
    \label{fig:pic_eef_idea}
  \end{figure*}

  Thus
  \begin{align}
    |\chi(\vec{r})|^2 := \braket{\psi(\vec{r},2,\dots,N)|\psi(\vec{r},2,\dots,N)}_{2 \dots N} \equiv \rho(\vec{r}) 
  \end{align}
  is the one-electron density normalized to 1 and $\braket{\dots}_{2 \dots N}$ indicates the scalar product (integral) with respect to the coordinates $\vec{r}_2, \dots, \vec{r}_N$.
  As the one-electron density is the marginal density of finding an electron at $\vec{r}$, independent of the location of the other electrons, $\chi(\vec{r})$ is called the marginal wavefunction.
  The function 
  \begin{align}
    \phi(2,\dots,N;\vec{r}) := \frac{\psi(\vec{r},2,\dots,N)}{\chi(\vec{r})}
    \label{eq:phi}
  \end{align}
  is the conditional wavefunction whose squared magnitude, $|\phi(2,\dots,N;\vec{r})|^2$,
  represents the conditional probability of finding electrons at 
  $\vec{r}_2, \dots, \vec{r}_N$, given an electron is located at $\vec{r}$.
  Thus, it has to obey the partial normalization condition
  \begin{align}
    \braket{\phi(2,\dots,N;\vec{r}) | \phi(2,\dots,N;\vec{r})}_{2 \dots N} \stackrel{!}{=} 1
    \label{eq:pnc}
  \end{align}
  for all values of $\vec{r}$.
  We call the electrons at $\vec{r}_2, \dots, \vec{r}_N$ the environment \cite{kocak2021}.
  The function $\phi$ encodes the spatial electron entanglement \cite{schroeder2017}, in the sense that the $N$-electron system is in general not the product of a one-electron wavefunction and an $(N-1)$-electron wavefunction, but that the wavefunction $\phi$ of the $N-1$ electrons depends on where the remaining electron of the $N$-electron system is found (measured). Expressing the above-mentioned potentials, and especially $v^{\rm G}$ in terms of $\phi$ allows us to give them a deeper, physical interpretation.
  From \eqref{eq:eef_ansatz} it follows that $\chi(\vec{r})$ obeys 
  the one-electron Schr\"odinger equation\cite{abedi2010,schild2017}
  \begin{align}
    \left( \frac{(-i \prv + \vec{A}(\vec{r}))^2}{2} +  v(\vec{r}) \right) \chi(\vec{r}) &= E \chi(\vec{r})
    \label{eq:eef_chi}
  \end{align}
  with the vector potential
  \begin{align}
    \vec{A}(\vec{r}) &= \braket{\phi(2,\dots,N;\vec{r}) |-i \prv \phi(2,\dots,N;\vec{r})}_{2 \dots N}
  \end{align}
  and with the scalar potential (cf.\ \cite{buijse1989,gritsenko1994,gritsenko1996a,gritsenko1996b})
  \begin{align}
    v(\vec{r}) &= v^{\rm T}(\vec{r}) + v^{\rm V}(\vec{r}) + v^{\rm G}(\vec{r}) + v^{\rm ext}(\vec{r})
    \label{eq:v_eef}
  \end{align}
  that consists of the terms
  \begin{align}
    v^{\rm T}(\vec{r})  
      &= \matrixel{\phi(2,\dots,N;\vec{r})}{-\sum\limits_{j=2}^N \frac{\nabla_j^2}{2}}{\phi(2,\dots,N;\vec{r})}_{2 \dots N} 
      \label{eq:eef_vt} \\
    v^{\rm V}(\vec{r})  
      &= \matrixel{\phi(2,\dots,N;\vec{r})}{V(1,2,\dots,N)}{\phi(2,\dots,N;\vec{r})}_{2 \dots N} - v^{\rm ext}(\vec{r})
      \label{eq:eef_vv} \\
    v^{\rm G}(\vec{r}) 
      &= \frac{1}{2} \left( \braket{\prv \phi(2,\dots,N;\vec{r}) | \prv \phi(2,\dots,N;\vec{r})}_{2 \dots N} - \left|\vec{A}(\vec{r}) \right|^2 \right).
      \label{eq:eef_vfs}
  \end{align}
  The term $v^{\rm T}(\vec{r})$ is the expectation value of the kinetic energy of the  $N-1$ environmental electrons and $v^{\rm V}(\vec{r})$ is the corresponding expectation value of the potential energy.
  It is useful for the discussion below to define the sum
  \begin{align}
    v^{\rm H}(\vec{r})   &= v^{\rm T}(\vec{r}) + v^{\rm V}(\vec{r}).   \label{eq:eef_vh}
  \end{align}  
  Hence $v^{\rm H}(\vec{r})$ is the expectation value of the energy of the environment given one additional electron is at $\vec{r}$.
  The geometric potential $v^{\rm G}(\vec{r})$ is the focus of our paper. It discussed in Sec. \ref{sec:geopot} and illustrated in Sec.\ \ref{sec:dia}.
  It is connected to how much the conditional wavefunction $\phi$ changes w.r.t.\ $\vec{r}$ and it is needed to calculate the correct kinetic energy of the electron in the presence of the electrons in the environment.
  Together with $\vec{A}(\vec{r})$, it describes the reaction of the environment to an infinitesimal change of the position of the additional electron at $\vec{r}$, and it is related to the Fubini-Study metric as well as to the quantum geometric tensor \cite{provost1980}.
  The meaning of the potentials is explained pictorially in Fig.\ \ref{fig:pic_eef_idea}.
  
  From the product form \eqref{eq:eef_ansatz} the one-electron wavefunction $\chi(\vec{r})$
  is defined only up to a phase, because we can replace $\chi(\vec{r})$ and $\phi(2,\dots,N;\vec{r})$ with
  \begin{subequations}
    \begin{align}
      \tilde{\chi}(\vec{r}) &:= e^{-i S(\vec{r})} \chi(\vec{r})  \\
      \tilde{\phi}(2,\dots,N;\vec{r}) &:= e^{+i S(\vec{r})} \phi(2,\dots,N;\vec{r}),
    \end{align}
    \label{eq:gauge}
  \end{subequations}
  where $S \in \mathbb{R}$, without changing the many-electron wavefunction $\psi(1,\dots,N)$ and without violating the partial normalization condition \eqref{eq:pnc}.
  The equations for $\chi$ (eq.\ \eqref{eq:eef_chi}) and $\phi$ (see the Supplemental 
  Material of \cite{schild2017}) also do not change under \eqref{eq:gauge} if $\vec{A}(\vec{r})$ is replaced with 
  \begin{align}
    \tilde{\vec{A}}(\vec{r}) = \vec{A}(\vec{r}) + \prv S(\vec{r}).
    \label{eq:Agauge}
  \end{align}
  The choice of $S$ is thus arbitrary, it is a gauge freedom of the theory.
  The measurable quantities of the theory need to be gauge invariant, i.e., they cannot depend on the choice of $S$.
  The potentials $v^{\rm T}$, $v^{\rm V}$, and $v^{\rm G}$ have this property, as shown in Appendix \ref{sec:eefpot}.
  Also, 
  \begin{align}
    \hat{\vec{p}} = -i \prv + \vec{A}(\vec{r})
  \end{align}
  is the gauge-invariant canonical momentum  and 
  \begin{align}
    \hat{T} = \frac{(-i \prv + \vec{A}(\vec{r}))^2}{2} + v^{\rm G}(\vec{r})
    \label{eq:ekin_gi}
  \end{align}
  is the gauge-invariant kinetic energy of an electron in the environment of the other electrons.
  
  An important feature of the EEF is that the many-electron problem is replaced with the one-electron problem \eqref{eq:eef_chi} for the one-electron wavefunction $\chi$.
  If the (components of the) potentials $v$ and $\vec{A}$ were known, one-electron observables of the many-electron system (like the dipole or the momentum) could be directly calculated from $\chi$ and the energy of the many-electron system $E$ could be obtained.
  However, to obtain these potentials the conditional wavefunction $\phi$ is needed, and its determining equation is difficult to solve exactly \cite{gossel2019}.
  Nevertheless, the EEF formalism provides explicit expressions for the needed one-electron potentials in terms of $\phi$ (or in terms of the full many-electron wavefunction $\psi$) that can be used to find suitable approximations. Furthermore, the dependence of the various potentials on $\phi$ provides valuable information, and offers a physical interpretation to these quantities both within the EEF formalism and within DFT. 
  This paves the way to advanced approximations to the exchange-correlation and Pauli terms.
  
  \subsection{Relation of the EEF and DFT}
    
  In this section, we show the relation of the EEF to DFT.
  In KS-DFT \cite{kohn1965}, the interacting many-electron system is replaced by the KS system, a non-interacting many-electron system with the same one-electron density as for the interacting problem.
  The wavefunction of the KS system (for a non-degenerate ground state) is then expressed as
  \begin{align}
    \psi^{\rm KS}(1,\dots,N)
      &= \frac{1}{\sqrt{N!}} \hat{A} \left( \prod_{j=1}^N \vp_j^{\rm KS}(j) \right) 
      \label{eq:psiks}
  \end{align}
  where $\hat{A}$ is an anti-symmetrization operator and $\vp_j^{\rm KS} (\vec{r})$ are the KS orbitals, which are obtained by solving the one-electron Schr\"odinger equation 
  \begin{align}
    \left( -\frac{\prv^2}{2} + v^{\rm KS}(\vec{r}) \right) \vp_j^{\rm KS}(\vec{r}) = \varepsilon_j^{\rm KS} \vp_j^{\rm KS}(\vec{r}).
  \end{align}
  The orbitals $\vp_j^{\rm KS} (\vec{r})$ are orthonormal, $\braket{\vp_i^{\rm KS} |\vp_j^{\rm KS} } = \delta_{ij}$ , and $\psi^{\rm KS}$ is normalized, $ \braket{\psi^{\rm KS}|\psi^{\rm KS}}= 1$.
  The KS potential,
  \begin{align}
    v^{\rm KS}(\vec{r}) = v^{\rm HXC}(\vec{r}) + v^{\rm ext}(\vec{r}),
  \end{align}
  is the sum of the external potential $v^{\rm ext} $ and the Hartree-exchange-correlation potential $v^{\rm HXC}$.
  The one-electron density of the interacting system is
  \begin{align}
    \rho(\vec{r}) 
      \equiv \frac{1}{N} \sum_{j=1}^N |\vp_j^{\rm KS}(\vec{r})|^2.
      \label{eq:norm}
  \end{align}
  In contrast to the usual convention of normalizing the one-electron density to the number of electrons, we require that $\braket{\rho(\vec{r})} = 1$, i.e., the many-electron wavefunction and the one-electron density are all normalized to 1.
  We make this non-standard choice for $\rho$ because we interpret the density $\rho(\vec{r})$ as the density of one electron in the environment of the other electrons.
  While the KS system exists and is unique \cite{hohenberg1964}, the KS potential $v^{\rm KS}$ cannot be directly obtained from the many-electron wavefunction $\psi$ and/or the one-electron density $\rho$.
  Different numerical methods exist to find the exact $v^{\rm KS}$ for a given one-electron density, and some recent discussions and applications  of this inverted KS problem can be found in \cite{hodgson2013,kananenka2013,hodgson2017,jensen2018,kumar2019,callow2020,kraisler2020,kraisler2020b}.
  
  The EEF equation \eqref{eq:eef_chi} is equivalent to the central equation 
  of OF-DFT,
  \begin{align}
    \left( -\frac{\prv^2}{2} + v^{\rm KS}(\vec{r}) + v^{\rm P}(\vec{r}) \right) \sqrt{\rho(\vec{r})} &= \mu \sqrt{\rho(\vec{r})}
    \label{eq:ofdft}
  \end{align}
  where $\mu = \varepsilon_N^{\rm KS}$ is the chemical potential (the eigenvalue of the highest occupied KS orbital) and $v^{\rm P}$ is the Pauli potential.
  From \eqref{eq:pauli} we see that \eqref{eq:ofdft} is identical to the EEF equation \eqref{eq:eef_chi} if we fix the gauge as $\vec{A}(\vec{r}) = 0$
  and if $\chi(\vec{r}) = \sqrt{\rho(\vec{r})}$, i.e., if $\chi$ is real-valued.
  This gauge choice cannot always be made \cite{requist2016,requist2017} but is supposed to be possible for the (non-degenerate) ground state of the many-electron system with zero total angular momentum and possibly also for other states without total angular momentum.\footnote{
    The gauge $\vec{A}(\vec{r}) = 0$ implies that the vector potential is curl-free in any gauge. 
    This condition may be violated even if the total angular momentum vanishes, hence we cannot make a definite statement here.}
  With this gauge, the EEF potential is related to KS and Pauli potentials as
  \begin{align}
    v(\vec{r}) &= v^{\rm KS}(\vec{r}) + v^{\rm P}(\vec{r}) \\
     v^{\rm H}(\vec{r}) + v^{\rm G}(\vec{r}) &= v^{\rm HXC}(\vec{r}) + v^{\rm P}(\vec{r})
  \end{align}
  up to a constant $E - \mu$ to be added on the right-hand side of the equations.
  
  The Pauli potential can be written in terms of the KS system as\cite{levy1988,kraisler2020}
  \begin{align}
    v^{\rm P}(\vec{r}) 
      = v^{\rm PH}(\vec{r}) + v^{\rm PG}(\vec{r}) 
    \label{eq:pauli_ks}
  \end{align}
  with 
  \begin{align}
    v^{\rm PH}(\vec{r}) 
      &= \sum_{n=1}^N (\varepsilon_{N}^{\rm KS} - \varepsilon_n^{\rm KS}) |\phi_n^{\rm KS}(\vec{r})|^2 
      \label{eq:ph} \\
    v^{\rm PG}(\vec{r}) 
      &= \frac{1}{2} \sum_{n=1}^N |\prv \phi_n^{\rm KS}(\vec{r})|^2,
      \label{eq:pg}
  \end{align}
  where we introduced the functions
  \begin{align}
    \phi_n^{\rm KS}(\vec{r}) = \frac{\vp_n^{\rm KS}(\vec{r})}{\sqrt{\rho(\vec{r})}}.
    \label{eq:phi_n_KS}
  \end{align}
  
  We now relate the EEF formalism to DFT by realizing that the functions \eqref{eq:phi_n_KS} are similar to the conditional wavefunction $\phi$ of the EEF and may be interpreted as KS orbitals of the environment.
  In particular, we define the conditional wavefunction 
  \begin{align}
    \phi^{\rm KS}(2,\dots,N;\vec{r}) = \frac{\psi^{\rm KS}(\vec{r},2,\dots,N)}{\sqrt{\rho(\vec{r})}}
    \label{eq:kscond}
  \end{align}
  of the non-interacting KS system, where $\psi^{\rm KS}$ is given by \eqref{eq:psiks}, and we can interpret the potential $v$ as a functional $v = v[\phi,V]$ of the conditional wavefunction $\phi$ and the many-electron potential $V(1,\dots,N)$, see \eqref{eq:v_eef}-\eqref{eq:eef_vfs}.
  As both $\psi$ and $\psi^{\rm KS}$ correspond to the same one-electron density $\rho(\vec{r})$, it follows that\cite{complet} 
  \begin{align}
    v[\phi,V] = v[\phi^{\rm KS},V^{\rm KS}], 
    \label{eq:v_vs_vni}
  \end{align}
  where $V^{\rm KS}(1,\dots,N) = \sum_{j=1}^N v^{\rm KS}(j)$ is the many-electron potential of the non-interacting KS system.
  Relation \eqref{eq:v_vs_vni} states that the same one-electron potential $v$ is obtained if it is evaluated as a functional of the exact quantities or of the KS quantities.
  
  If we also interpret $v^{\rm H}$ and $v^{\rm G}$ as functionals $v^{\rm H} = v^{\rm H}[\phi,V]$ and $v^{\rm G} = v^{\rm G}[\phi]$ and evaluate them with the KS quantities, we find that (cf.\ \cite{gritsenko1996b})
  \begin{subequations}
    \begin{align}
      v^{\rm H}[\phi^{\rm KS},V^{\rm KS}](\vec{r}) 
      &=v^{\rm PH}(\vec{r}) + v^{\rm HXC}(\vec{r}) \label{eq:vh_corr}  \\
      v^{\rm G}[\phi^{\rm KS}](\vec{r})
      &=v^{\rm PG}(\vec{r}) , \label{eq:vg_corr}
    \end{align}
    \label{eq:v_corr}
  \end{subequations}
  where \eqref{eq:vh_corr} holds up to a constant $(E-\mu)$, as explained above.
  We thus see that $v^{\rm PG}(\vec{r})$ is the geometric potential of the $N-1$ non-interacting electrons of the KS system if one additional electron is at $\vec{r}$.
  Also, $v^{\rm H}[\phi^{\rm KS},V^{\rm KS}]$ is the corresponding energy of those $N-1$ electrons. 
  The left-hand side and the right-hand side of \eqref{eq:vh_corr} are only equal up to a constant (which is $ E - \epsilon_N^{\rm KS}$), because $v$ in \eqref{eq:eef_chi} and $v^{\rm KS} + v^{\rm P}$ in \eqref{eq:ofdft} are shifted relative to each other.
  In the EEF, the asymptotic value  $\lim_{|\vec{r}| \rightarrow \infty} v(\vec{r})$ is the energy of the ionized system, while in OF-DFT the potential typically becomes zero for large $|\vec{r}|$.
  
  \subsection{DFT potentials interpreted via the EEF}
  
  Via  Eqs. \eqref{eq:vh_corr} and \eqref{eq:vg_corr}, the EEF provides a different view on the Hartree-exchange-correlation potential $v^{\rm HXC}$ and, in particular, on the Pauli potential $v^{\rm P}$ compared to the usual interpretation:
  When OF-DFT and KS-DFT are compared, a central point of discussion is how the two theories treat the fermionic antisymmetry of a many-electron system.
  The symmetry constraints for the many-electron wavefunction $\psi(1,\dots,N)$ are included in an elegant way in KS-DFT via the construction of the non-interacting KS system, which has the same one-electron density like the interacting system, but which also corresponds to an antisymmetric many-electron wavefunction $\psi^{\rm KS}(1,\dots,N)$.
  OF-DFT, however, is sometimes interpreted as mapping to a non-interacting bosonic system with the same one-body density.
  The Pauli potential is thus often viewed as necessary to describe the antisymmetry correctly, because it is the difference potential between the supposed fermionic and bosonic non-interacting systems. \cite{march1986,levy1988}.
  
  While the construction of the non-interacting bosonic system is technically correct, the EEF provides a different interpretation, which sheds light on the nature of the Pauli potential:
  Despite the product form \eqref{eq:eef_ansatz}, the fermionic antisymmetry constraints are unbroken:
  The wavefunction $\phi$ fulfills the symmetry constraints w.r.t.\ exchange of the coordinates of the electrons in the environment $\vec{r}_2, \dots, \vec{r}_N$, i.e., if space and spin coordinates are exchanged,
  \begin{align}
    \phi(2,\dots, j, \dots, k, \dots, N;1) = -\phi(2,\dots, k, \dots, j, \dots, N;1).
  \end{align}
  The antisymmetry constraints w.r.t.\ the additional electron at $\vec{r}$ are found in the product $\chi(1) \phi(2,\dots,N;1)$, e.g.
  \begin{align}
    \chi(1) \phi(2,\dots,j,\dots,N;1) = -\chi(j) \phi(2,\dots,1,\dots,N;j),
  \end{align}
  and are thus implicitly contained in the EEF formalism.
  
  In the EEF picture, there is thus no Pauli potential which turns a (non-interacting) bosonic system into a  fermionic system, but the interacting fermionic system itself is considered from the start.
  The EEF potentials $v^{\rm H}$ and $v^{\rm G}$ have a clear physical meaning in terms of how one electron feels the environment provided by the other electrons:
  $v^{\rm H}$ is the energy of the other electrons and $v^{\rm G}$ is an additional resistance that the electron experiences if its change of position leads to a change of the state of the other electrons (i.e., if there is a strong spatial entanglement).
  
  If the interacting system is replaced with the KS system, we see from \eqref{eq:v_corr} that the corresponding geometric potential becomes one part of the Pauli potential $v^{\rm PG}$, while the energy of the environment becomes the sum of $v^{\rm HXC}$ with the other part of the Pauli potential $v^{\rm PH}$.
  Moreover, by evaluating \eqref{eq:vh_corr} explicitly, we have (up to a constant)
  \begin{align}
    v^{\rm H}[\phi^{\rm KS},V^{\rm KS}] + v^{\rm ext}
      &= \Braket{\phi^{\rm KS} | -\sum_{j=2}^N \frac{\nabla_j^2}{2} + \sum_{j=1}^N v^{\rm KS}(j) | \phi^{\rm KS}}_{2 \dots N} \\
      &= v^{\rm PH} + \underbrace{\Braket{\phi^{\rm KS} | v^{\rm KS}(1) | \phi^{\rm KS}}_{2 \dots N}}_{v^{\rm HXC} + v^{\rm ext}}.
  \end{align}
  From this relation and from \eqref{eq:v_corr}, we can interpret the Pauli potential as the EEF potential of the KS system, where its two parts are the energy of the environment $v^{\rm PH}$ and the geometric potential $v^{\rm PG}$.
  The external potential $v^{\rm ext}$ of the interacting system, however, is replaced for the KS system by $v^{\rm ext} + v^{\rm HXC}$.
  
  In contrast to the usual view on DFT and especially on OF-DFT, in the EEF perspective, one does not talk about a fermionic or a bosonic many-electron problem.
  Instead, the problem of one electron in the environment of other electrons is considered both for the interacting many-electron problem (the EEF) and for the KS system (OF-DFT).
  In both cases the same one-electron potential $v$ is obtained, but because the environment is described differently, the contributions to $v$ can differ, as well.
  
  \section{The geometric potential}
  \label{sec:geopot}
  
  While the average energy $v^{\rm H}$ of the environment is straightforward to understand in the EEF, the meaning of the geometric potential $v^{\rm G}$ is less obvious.
  As \eqref{eq:ekin_gi} shows, $v^{\rm G}$ is one part of the gauge-invariant kinetic energy operator $\hat{T}$ and $\braket{\chi|\hat{T}|\chi}$ is the expectation value of the kinetic energy of one electron in the environment of the other $N-1$ electrons.
  Furthermore, in Appendix \ref{sec:ekind} we show that $v^{\rm G}$ is related to the positively-defined kinetic energy density (cf.\ \cite{buijse1989}) and that the different kinetic energy densities of the interacting and the KS system fully account for the differences between $v^{\rm G}$ and $v^{\rm PG}$.
  Hence, $v^{\rm G}$ has been called the ``kinetic potential'' in the literature \cite{gritsenko1994,giarrusso2020}.
  
  However, there is a geometric meaning attached to $v^{\rm G}$ and, in this context, also to the vector potential $\vec{A}$.
  In this section, we first show how $v^{\rm G}$ and $\vec{A}$ are related to the quantum-geometric tensor that describes the geometric structure of the environment of the one-electron system.
  We then proceed by examining the effect of scaling and translation of the wavefunction $\phi$ of the environment, and we analyze the geometric picture if $\phi$ can be represented by a two-state model.
  
  \subsection{The quantum-geometric tensor}
  \label{sec:qgt}
  
  To better understand $v^{\rm G}$, we describe its relation to the quantum geometric tensor.
  We consider a general function $f(x;t)$ that is an element of a Hilbert space with inner product defined w.r.t.\ the coordinate(s) $x$.
  The function $f$ has an additional dependence on a parameter $t$, and it shall also be normalized as $\braket{f(x;t)|f(x;t)}_x = 1$.
  We are interested in the change of $f$ with $t$.
  Taking the norm of the difference between $f(x;t+dt)$ and $f(x;t)$,
  \begin{align}
    df^2 = || f(x;t+dt) - f(x;t) ||^2 = \Braket{f(x;t+dt) - f(x;t) | f(x;t+dt) - f(x;t)}_x
  \end{align}
 we find for infinitesimal $dt$ that
 \begin{align}
   df^2 = \Braket{\pt f(x;t) | \pt f(x;t)}_x dt^2.
 \end{align}
 The term
 \begin{align}
    \tilde{g} = \Braket{\pt f(x;t) | \pt f(x;t)}_x \ge 0
    \label{eq:g01}
 \end{align}
 looks like a metric that represents how $f$ changes with $t$.
 
\begin{figure}[!htbp]
    \centering
    \includegraphics[width=0.5\textwidth]{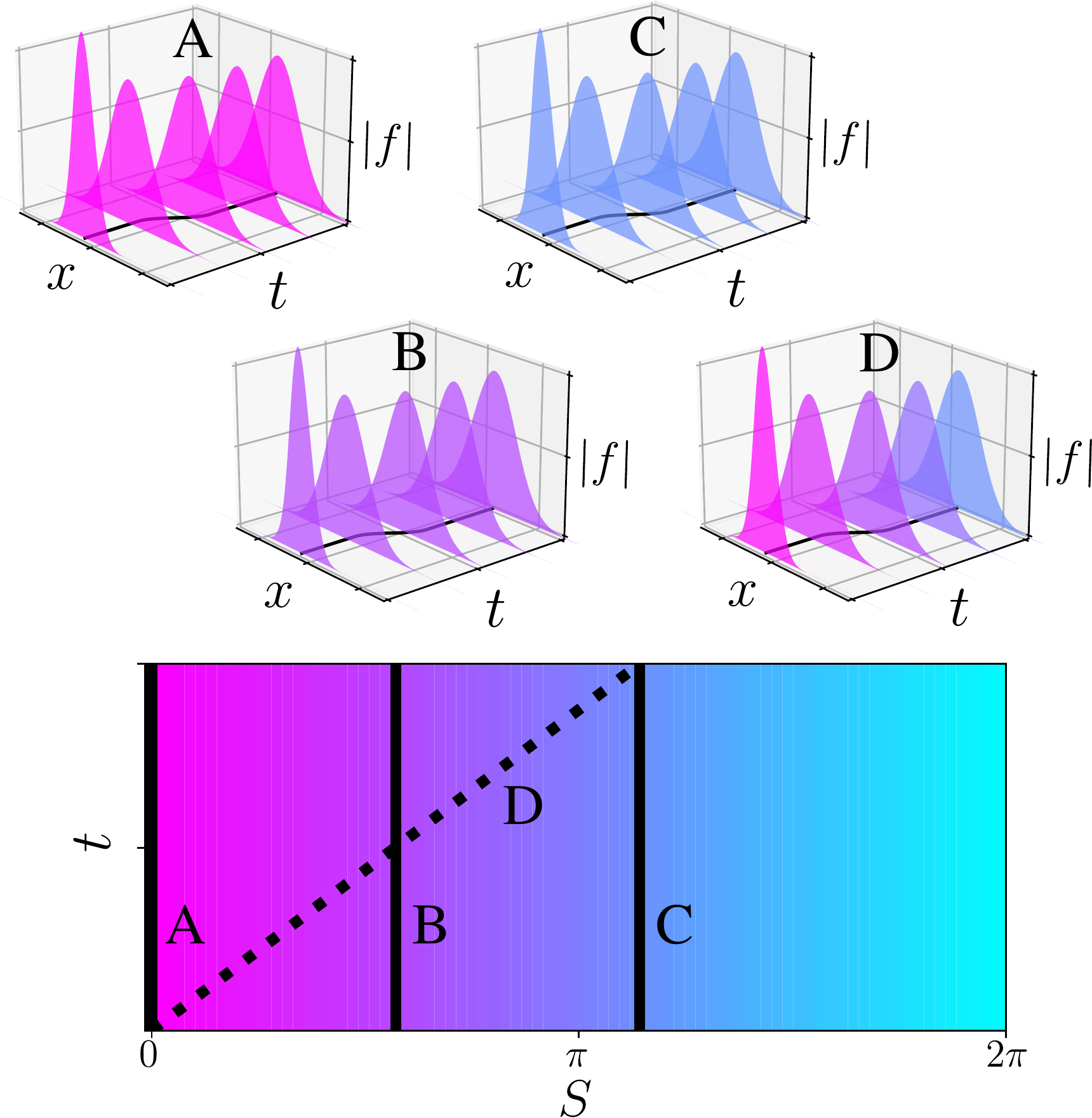}
    \caption{
    The Fubini-Study metric $g(t)$ measures the change of a function $f(x;t)$ with $t$, independent of the choice of some additional arbitrary phase $S(t)$.
    In the upper part of the figure, the plots A-D show the magnitude of some $f(x;t)$ for five different values of $t$.
    We have that $f(x;t) = |f(x;t)| e^{i \alpha(x;t)} e^{i S(t)}$, where $S(t)$ is arbitrary.
    The colors of the functions in plots A-D indicate the value of $S(t)$, which is given in the lower part of the figure.
    We require $g(t)$ to be the same for all choices of $S(t)$, i.e., for all paths A-D.
    This is what the projector $P_{\perp}$ defined in \eqref{eq:proj} achieves.
    }
    \label{fig:proj}
  \end{figure}

 However, \eqref{eq:g01} is ambiguous, as $f(x;t)$ is only defined up to an $x$-independent phase.
 Indeed, the transformation
 \begin{align}
   f(x;t) \rightarrow f(x;t) e^{-i S(t)}
 \end{align}
 with $S \in \mathbb{R}$, does not change the state that $f$ represents, but does affect $\tilde{g}$, as $\pt f$ changes with $S$ as
 \begin{align}
   \pt f(x;t) \rightarrow e^{-i S(t)} \pt f(x;t) - i \pt S e^{-i S(t)} f(x;t) \, .
 \end{align} 
 The choice of $S(t)$ is a gauge choice and, for $\tilde{g}$ to be a useful metric, $\tilde{g}$ should be independent of the choice of gauge. As the gauge-dependent term of $\pt f$ depends linearly on $f(x;t)$, gauge independence is achieved by projecting out this term with a projector $P_{\perp}$ into the space orthogonal to $f$
   \begin{align}
    P_{\perp} = 1 - \ket{f(x;t)}_x \bra{f(x;t)},
    \label{eq:proj}
  \end{align}
  with $P_{\perp} \ket{f} = 0$.
  Using the projector $P_{\perp}$ in \eqref{eq:g01}, we obtain a gauge independent metric, the Fubini-Study metric \cite{provost1980,abe1993}
  \begin{align}
    g = \Braket{\pt f(x;t) | P_{\perp} |  \pt f(x;t)}_x \ge 0 \, .
    \label{eq:g02}
  \end{align}
  The problem that $P_{\perp}$ solves is represented graphically in Fig.\ \ref{fig:proj}.

  If $f$ depends on multiple parameters, $f(x;t_1,\dots,t_n)$, the change of $f$ with the parameters is described by the quantum-geometric tensor \cite{provost1980,berry1989}
  \begin{align}
    T_{ij} = \Braket{\partial_{t_i} f(x;t_1,\dots,t_n) | P_{\perp} | \partial_{t_j} f(x;t_1,\dots,t_n)}.
  \end{align}
  This tensor yields the Fubini-Study metric tensor as its real (symmetric) part,
  \begin{align}
    g_{ij} = \frac{1}{2}(T_{ij} + T_{ji}), 
  \end{align}
  and its imaginary (antisymmetric) part is the Berry curvature \cite{berry1984}
  \begin{align}
    B_{ij} = \frac{1}{2i}(T_{ij} - T_{ji}).
  \end{align}
  
  We can compare these definitions to the geometric potential $v^{\rm G}$.
  There, $f(x;t) \rightarrow \phi(2,\dots,N;\vec{r})$ where the coordinates $\vec{r}$ of one electron are parameters.
  As we restrict the discussion to a simple kinetic energy operator \eqref{eq:se} with Cartesian coordinates for the particles, there are no terms coupling the three components $r_i$ of $\vec{r}$ and, thus, the corresponding Fubini-Study metric tensor is diagonal with components
  \begin{align}
    g_{ii} = \Braket{ \partial_{r_i} \phi(2,\dots,N;\vec{r}) | P_{\perp} | \partial_{r_i} \phi(2,\dots,N;\vec{r})}_{2 \dots N}.
  \end{align}
  The metric is 
  \begin{align}
    ds^2 = \sum_i \sum_j g_{ij} dr_i dr_j = \sum_{i=1}^3 g_{ii} dr_i^2
    \label{eq:ds01}
  \end{align}
  and the geometric potential is
  \begin{align}
   v^{\rm G}(\vec{r}) = \frac{\hbar^2}{2 m_{\rm e}} \sum_{i=1}^3 g_{ii},
   \label{eq:gpotii}
  \end{align}
  where we added Plank's constant and the electronic mass for emphasis.
  In atomic units and for the simple kinetic energy term in \eqref{eq:se}, the geometric potential is 
  \begin{align}
    v^{\rm G}(\vec{r}) = \frac{1}{2} \sum_{i=1}^3 g_{ii}.
  \end{align}
  The geometric potential $v^{\rm G}$ thus measures how much the wavefunction $\phi$ of the electrons in the environment changes when the position $\vec{r}$ of the conditional electron is changed.
  As it is a distance measure, $v^{\rm G}(\vec{r}) \ge 0$, which can be seen from its definition.
  Interpreted as a potential, it repels the electron from regions where the environment changes significantly along $\vec{r}$.
  
  Next to the metric tensor, the Berry curvature also contains important information about the geometry of the problem.
  In particular, its gauge-invariant components can be expressed in terms of the components of the vector potential $\vec{A}(\vec{r})$ as 
  \begin{align}
    B_{ij} = \partial_{r_i} A_j - \partial_{r_j} A_i.
  \end{align}
  It can be shown (e.g.\ by noting that $B_{ij}$ is the exterior derivative of $\vec{A}$ and by using the theory of differential forms \cite{flanders1989}) that, if any component $B_{ij} \ne 0$, the vector potential $\vec{A}(\vec{r})$ cannot be written as gradient of a scalar field, $\vec{A}(\vec{r}) \ne \nabla_{\vec{r}} S(\vec{r})$ for any $S \in \mathbb{R}$.
  In this case, we see from a comparison to \eqref{eq:Agauge} that the choice of gauge $\vec{A}(\vec{r}) \stackrel{!}{=} 0$ cannot be made.
  
  Below, we only consider (finite) one-dimensional systems for which $\vec{A}$ is a scalar field that can always be written as a gradient field.
  Then, the choice $A \stackrel{!}{=} 0 $ is possible.
  We make this choice and do not consider the vector potential or Berry curvature further.
  However, these quantities might have to be taken into account for the study of the full three-dimensional problem.
  Current results suggest that the choice $\vec{A}(\vec{r}) \stackrel{!}{=} 0$ is possible for the three-dimensional problem when the system is not rotating \cite{requist2016}, but further investigations are needed to draw a definite conclusion.

   In case of one-dimensional systems, the geometric potential reduces to
  \begin{align}
    v^{\rm G}(x_1) = \frac{1}{2} g_{11} = \frac{1}{2} \Braket{ \partial_1 \phi| \left( 1 - \Ket{\phi}\Bra{\phi} \right) | \partial_1 \phi}_{2}, \label{eq:g11}
  \end{align}
  and then the Fubini-Study distance $ds$ relates to the scaled distance $dx_1$ as
  \begin{align}
    ds = \sqrt{2v^{\rm G}(x_1)} \, dx_1 .
  \end{align}
  
  \subsection{Translation of a state} \label{sec:tra}

  To learn more about the effect of a change of the environment on $v^{\rm G}$, we consider various one-dimensional systems. 
  Each system is characterized by a conditional wavefunction $\phi (x_2; x_1)$, and a change of the shape of $\phi(x_2;x_1)$ is reflected in $v^{\rm G}$.
  
\begin{figure}[!htbp]
    \centering
    \includegraphics[width=.5\linewidth]{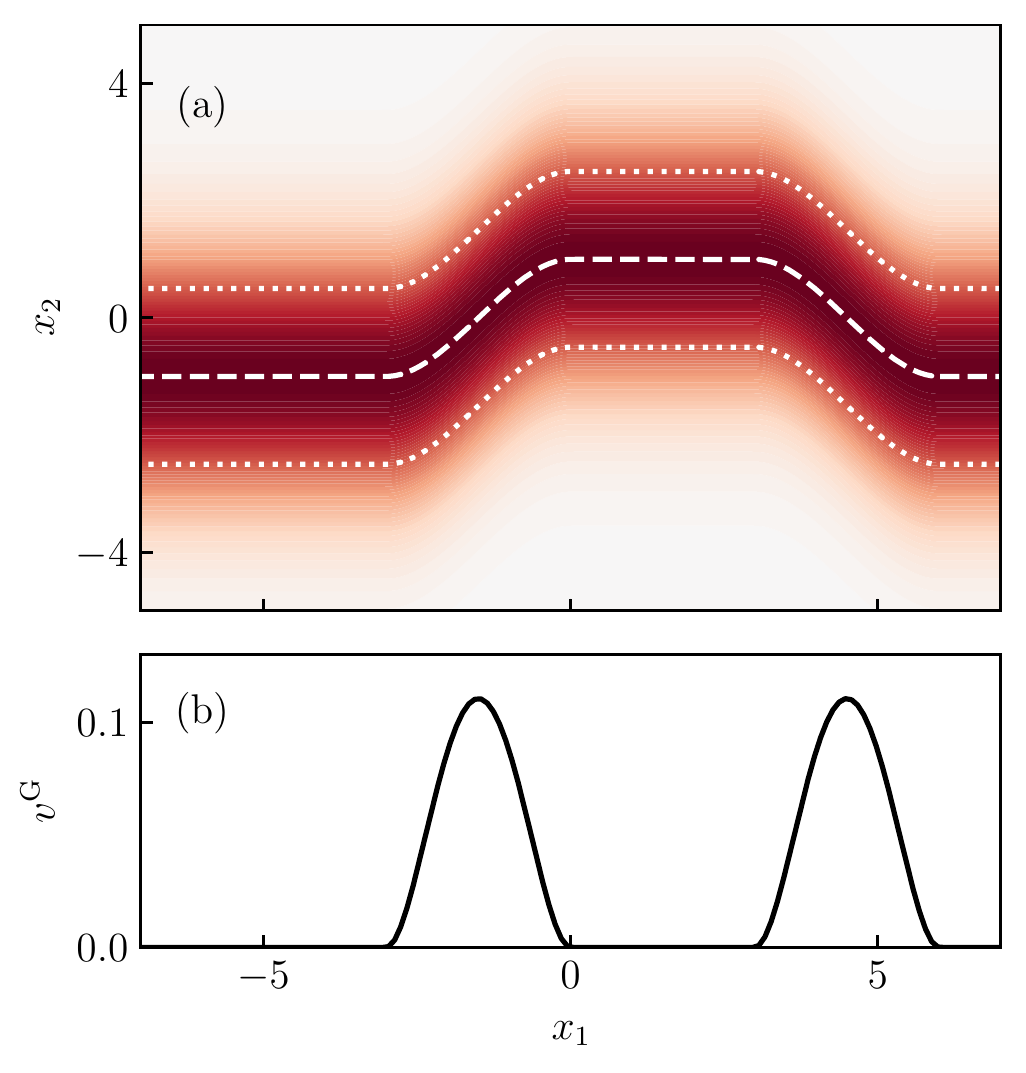}
    \caption{(a) The conditional wavefunction $\phi(x_2|x_1)$ and (b) the corresponding geometric potential $v^{\rm G}$ for a translation of a Gaussian function with marked center and width.
    The more rapid $\phi(x_2|x_1)$ is tanslated along $x_2$ when changing $x_1$, the higher $v^{\rm G}$ becomes.}
    \label{fig:trans}
  \end{figure}

  First, we look at a wavefunction $\phi (x_2; x_1)$ that does not change the shape with respect to $x_1$, but the state is translated along $x_2$.
  Such wavefunction is described as
  \begin{align}
  	\phi(x_2;x_1) = \phi_0(x_2 - a (x_1)) \, ,
  \end{align}
  where $\phi_0$ is a quantum state and $a (x_1)$ is a function describing the motion of the center (or, actually, any point) with respect to $x_1$ (see Fig.\ \ref{fig:trans}a). 
  From \eqref{eq:g11}, we get the Fubini-Study metric
  \begin{align}
  	g_{11} (x_1) = \left( \frac{d a}{ dx_1} \right)^2 \sigma_p^2 \, , \label{eq:g11trans}
  \end{align}
  where $\sigma_p^2$ is the uncertainty of the momentum $p$ of the state $\phi_0$, i.e.
  \begin{align}
    	\sigma_p^2 = \braket{ \delta \hat p^2}  = \braket{ \hat p^2} - \braket{\hat p}^2 \, ,
  \end{align}
  where $\delta \hat p := \hat p - \braket{\hat p}$ is the deviation operator of the momentum operator $\hat p = -i \partial_x$ and $\braket{\cdot}$ is the expectation value for the quantum state $\phi_0(x)$.

  By definition, the uncertainty $\sigma_p^2$ can only have non-negative values. 
  Furthermore, it equals to zero if and only if the state $\phi_0$ is an improper eigenstate $e^{i p x_2}$ of the operator $\hat p$ with the eigenvalue $p$. 
  For any normalizable state it is strictly positive.
  
  From \eqref{eq:g11trans} and \eqref{eq:gpotii} it is clear that the geometric potential $v^{\rm G}$ (that is just $g_{11}$ multiplied by a constant) is proportional to the square of ``the speed'' along the trajectory $a (x_1)$, i.e., to how much the state is translated along $x_2$ with changing $x_1$.
  For a slow translation we obtain small values of the geometric potential, whereas any rapid translation along $x_2$ with changing $x_1$ shows as a large peak in $v^{\rm G}$, as is shown in Fig.\ \ref{fig:trans}b.

  As the system is restrained to one dimension, the traversed Fubini-Study distance $s$ can be obtained from the integral of $\sqrt{2v^{\rm G}} $ as
  \begin{align}
  	\int ds = \int^{x_1 }  \sqrt{2v^{\rm G}(x_1')} \, dx_1' = \sigma_p \int^{x_1 }  \left| \frac{d a}{ dx_1} \right| \, dx_1' =  \sigma_p R(x_1) \, 
  \end{align}
  where $R(x_1)$ measures the traversed distance along the trajectory $a(x_1)$. The Fubini-Study distance is then the traversed distance in units of $1/\sigma_p$.

  We also note that there is a degree of freedom in the choice of the translation function $a \rightarrow a + k$, which does not affect the Fubini-Study distance or the geometric potential.
  
  \subsection{Scaling of a state} \label{sec:sca}
  
\begin{figure}[!htbp]
    \centering
    \includegraphics[width=.5\linewidth]{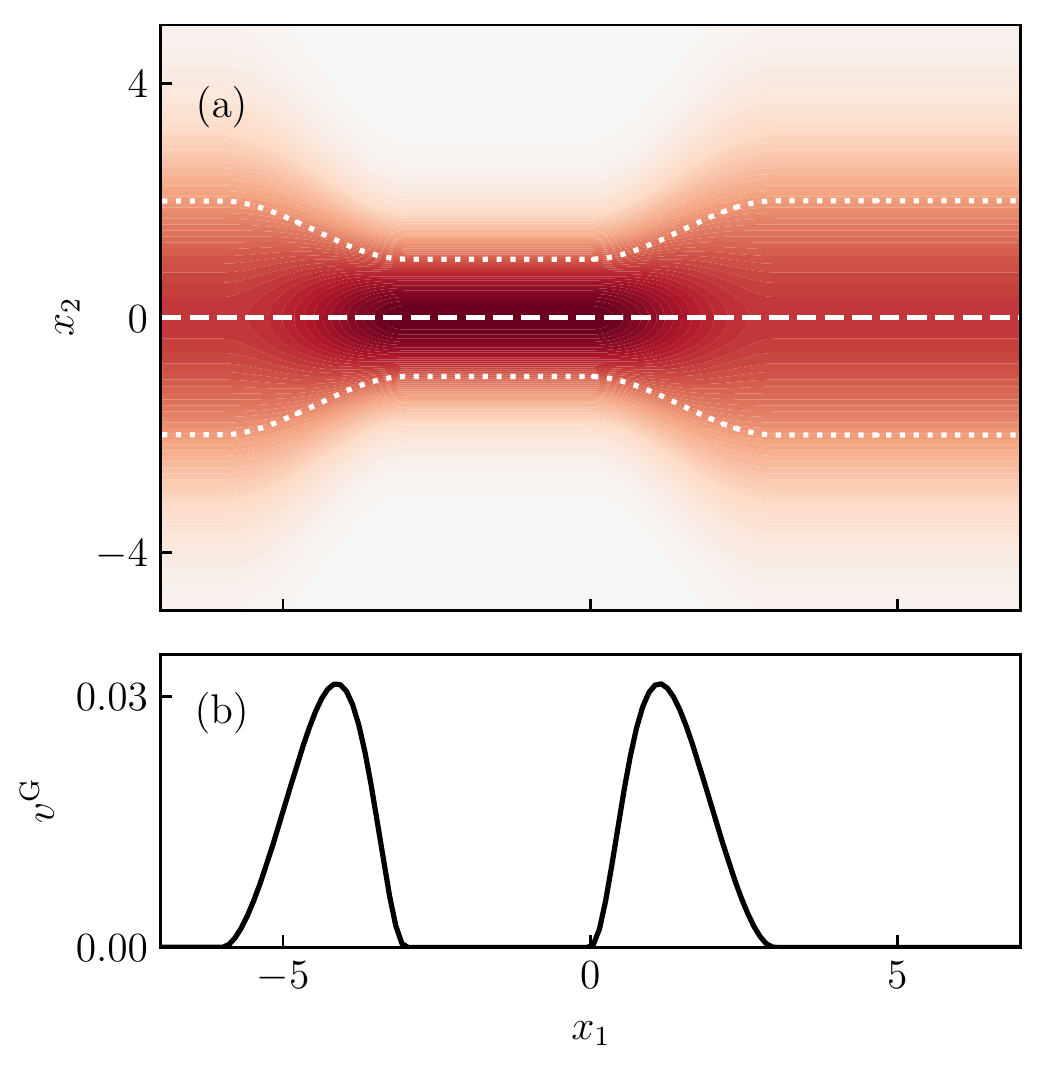}
    \caption{(a) The conditional wavefunction $\phi(x_2|x_1)$ and (b) the corresponding geometric potential $v^{\rm G}$ for a scaling of a Gaussian function with marked center and width.
    The more rapid the width of $\phi(x_2|x_1)$ changes when changing $x_1$, the higher $v^{\rm G}$ becomes.}
    \label{fig:scale}
  \end{figure}

  The next transformation that we consider is the scaling that broadens or compresses the shape of the quantum state. 
  Such a transformation is described as
  \begin{align}
  	\phi(x_2;x_1) = \frac{1}{\sqrt{\Delta(x_1)}} \phi_0 \left( \frac{x_2}{\Delta(x_1)} \right) \, ,
  \end{align}
  where $\phi_0$ is a quantum state and $\Delta(x_1)$ is a function describing the scaling with respect to $x_1$ (see Fig.\ \ref{fig:scale}).  
  The prefactor in front of $\phi_0$ is due to the normalization.
  The Fubini-Study metric is then
  \begin{align}
  	g_{11} (x_1) = \left( \frac{d}{ dx_1} \ln \Delta \right)^2 \sigma_b^2 \, , \label{eq:g11scale}
  \end{align}
  where $\sigma_b^2$ is the uncertainty of the operator $\hat b:= \frac{1}{2} (\hat x \hat p + \hat p \hat x)$ of the state $\phi_0$, i.e.
  \begin{align}
  	\sigma_b^2 = \braket{ \delta \hat b^2 } = \braket{ \hat b^2 } - \braket{\hat b }^2 \, ,
  \end{align}
  where $\delta \hat b := \hat b - \braket{\hat b}$ is the deviation operator of $\hat b$.
  
  The uncertainty $\sigma_b$ is a non-negative value reaching zero if and only if the state $\phi_0$ is an improper eigenstate $(\pm x)^{-\frac{1}{2} \pm i b}$ of the operator $\hat b$ with the eigenvalue $\pm b$. Therefore, for any normalizable state it is strictly positive.
  
  Similarly to the translation, we see from \eqref{eq:g11scale} and \eqref{eq:gpotii} that the geometric potential $v^{\rm G}$ is proportional to the square of ``the speed'' of the function $\ln \Delta$ with respect to $x_1$. Likewise, any slow or rapid change in the function manifests as a small or large peak in the geometric potential, respectively.
  
  The traversed Fubini-Study distance $s$ can be again obtained from the integral of $\sqrt{2v^{\rm G}} $ as
  \begin{align}
  	\int ds = \int^{x_1 }  \sqrt{2v^{\rm G}(x_1')} \, dx_1' = \sigma_b \int^{x_1 }  \left| \frac{d }{ dx_1} \ln \Delta \right| \, dx_1' =  \sigma_b D(x_1) \, 
  \end{align}
  where $D(x_1)$ measures the traversed distance in $\ln \Delta(x_1)$ in units of  $1/\sigma_b$.

  We also note that there is a degree of freedom in the choice of the scaling function $\Delta \rightarrow c \Delta$, which does not affect the Fubini-Study distance or the geometric potential.
  
  \subsection{Translation and scaling of a state}
  
\begin{figure}[!htbp]
    \centering
    \includegraphics[width=.5\linewidth]{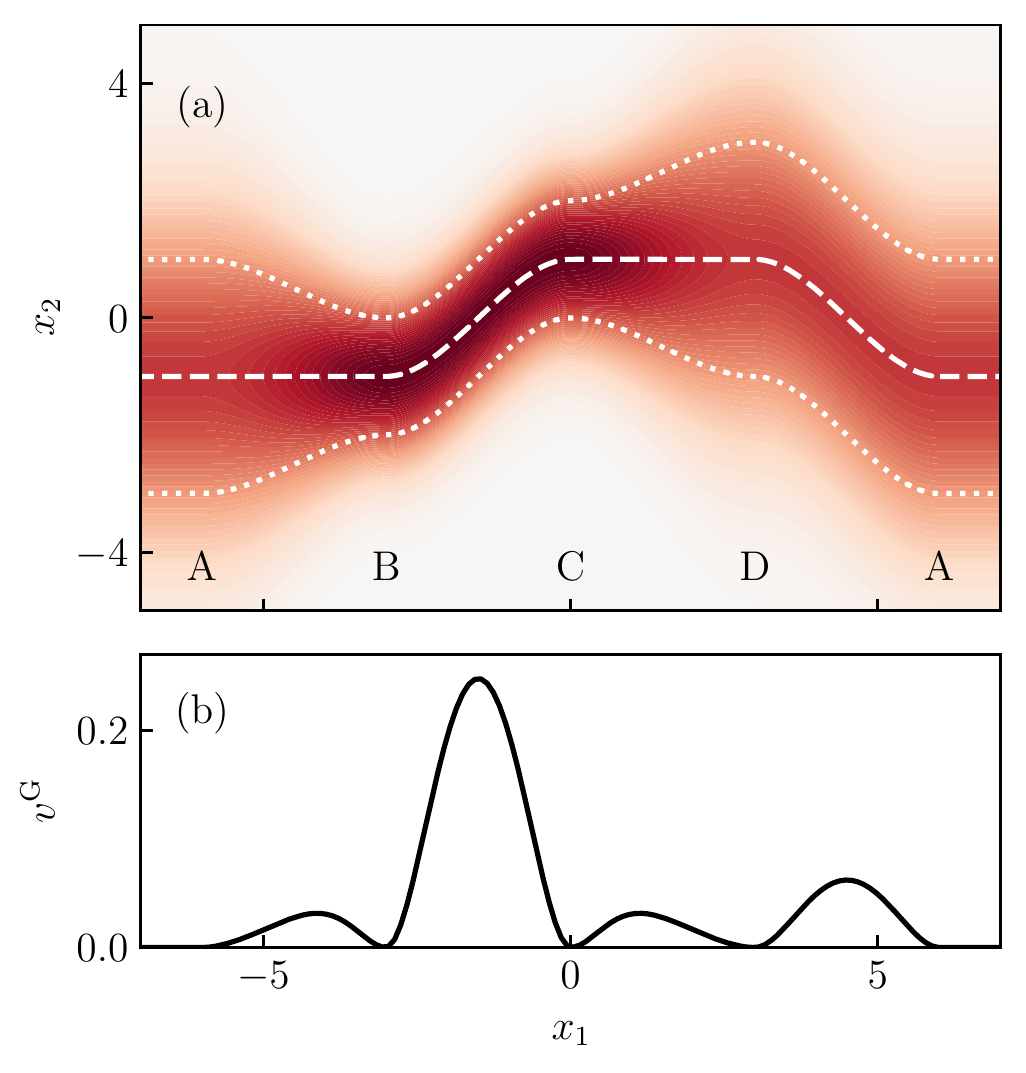}
    \caption{(a) The conditional wavefunction $\phi(x_2|x_1)$ and (b) the corresponding geometric potential $v^{\rm G}$ for combination of translation and scaling of a Gaussian function with marked center and width of the Gaussian function.
    As $\phi(x_2|x_1)$ is squezzed along $x_2$ when the rapid translation around $x_1$ happens, the peak in the geometric potential is higher than the corresponding peak in Fig.\ \ref{fig:trans}b for translation alone.}
    \label{fig:comb}
  \end{figure}

  Last, we combine translation and scaling, which results in the conditional wavefunction
 \begin{align}
	\phi(x_2;x_1) = \frac{1}{\sqrt{\Delta(x_1)}} \phi_0 \left( \frac{x_2}{\Delta(x_1)} - a (x_1) \right)  \, ,
 \end{align}
  where $\Delta$ and $a$ are two parametric functions (see Fig.\ \ref{fig:comb}). Thus, the Fubini-Study metric $g_{11}$ in one dimension can be interpreted as part of a two-dimensional Fubini-Study metric depending on the parameters $\Delta$ and $a$,
  
  \begin{align}
  	g_{11} (x_1) =
	\begin{pmatrix}
	\frac{d \Delta}{d x_1 } & \frac{d a}{d x_1 }
	\end{pmatrix}
	\begin{pmatrix}
	g_{\Delta \Delta} & g_{\Delta a} \\
    g_{a \Delta} & g_{a a}
    \end{pmatrix}
	\begin{pmatrix}
	\frac{d \Delta}{d x_1 } \\
	\frac{d a}{dx_1 }
	\end{pmatrix} \, ,
  \end{align}
  where the elements of the metric tensor are
  \begin{align}
    g_{\Delta \Delta} &= \frac{1}{\Delta^2} \braket{\delta \hat b^2 } \, , \\	
    g_{\Delta a} = g_{a \Delta} &= \frac{1}{\Delta^2} \Re \braket{\delta \hat b \delta \hat p } \, , \\
    g_{a a} &= \frac{1}{\Delta^2} \braket{\delta \hat p^2 } \, .
  \end{align}
  As in previous models, there is a degree of freedom in the chosen functions $\phi_0, \Delta$ and $a$. By appropriate translation of the state $\phi_0(x_2) \rightarrow \phi_0(x_2 - k)$, the off-diagonal terms can be set to zero,
  \begin{align}
    \frac{1}{\Delta^2}  \Re \braket{\delta \hat b \delta \hat p } &\rightarrow  \frac{1}{\Delta^2}  \Re \braket{\delta \hat b \delta \hat p } + k  \frac{1}{\Delta^2}  \braket{ \delta \hat p^2 } \, , \\
    k &= - \frac{ \Re \braket{\delta \hat b \delta \hat p } }{\braket{ \delta \hat p^2 }} \, .
  \end{align}
  After this translation, the final Fubini-Study metric is
  \begin{align}
    ds^2 = \sigma_b^2 (d \ln \Delta)^2 + \frac{\sigma_p^2}{\Delta^2} d a^2 \, . \label{eq:ds2-combi}
  \end{align}
  For $\Delta(x_1) = 1$, we recover the result for the translation (Eq.\ \eqref{eq:g11trans}), and, for $a(x_1) = \mathrm{const.}$, we recover the result for the scaling (Eq.\ \eqref{eq:g11scale}).


  \begin{figure}[!htbp]
    \centering
    \begin{overpic}[trim={0cm 0cm 0cm 0cm},clip,width=0.4\linewidth]{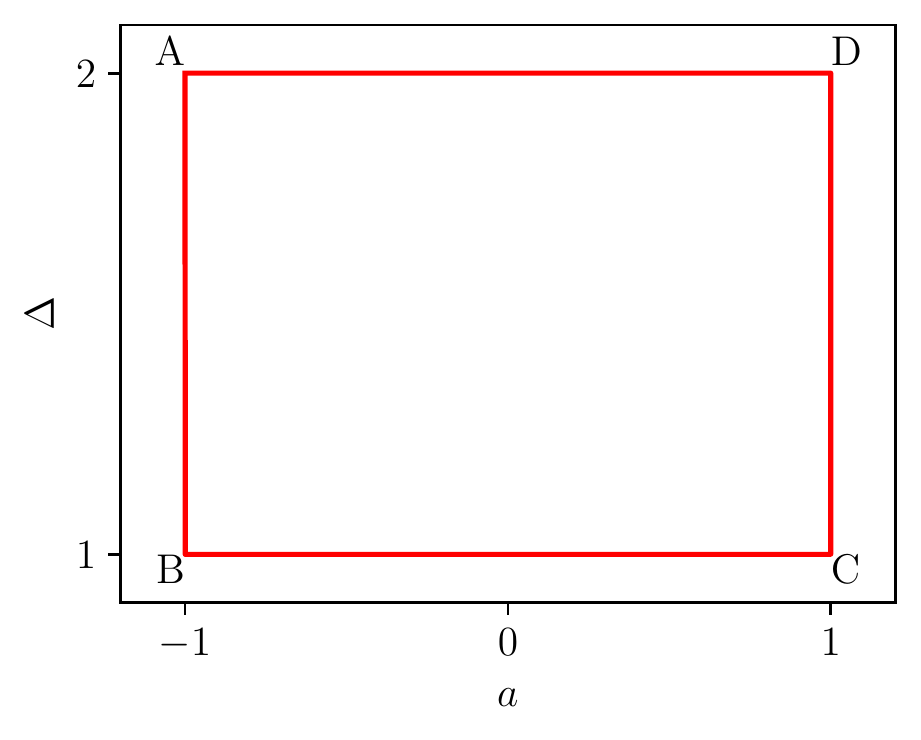}
    \put (20,83) {$(a)$}
    \end{overpic}
    \begin{overpic}[trim={0cm 0cm 0cm 0cm},clip,width=0.5\linewidth]{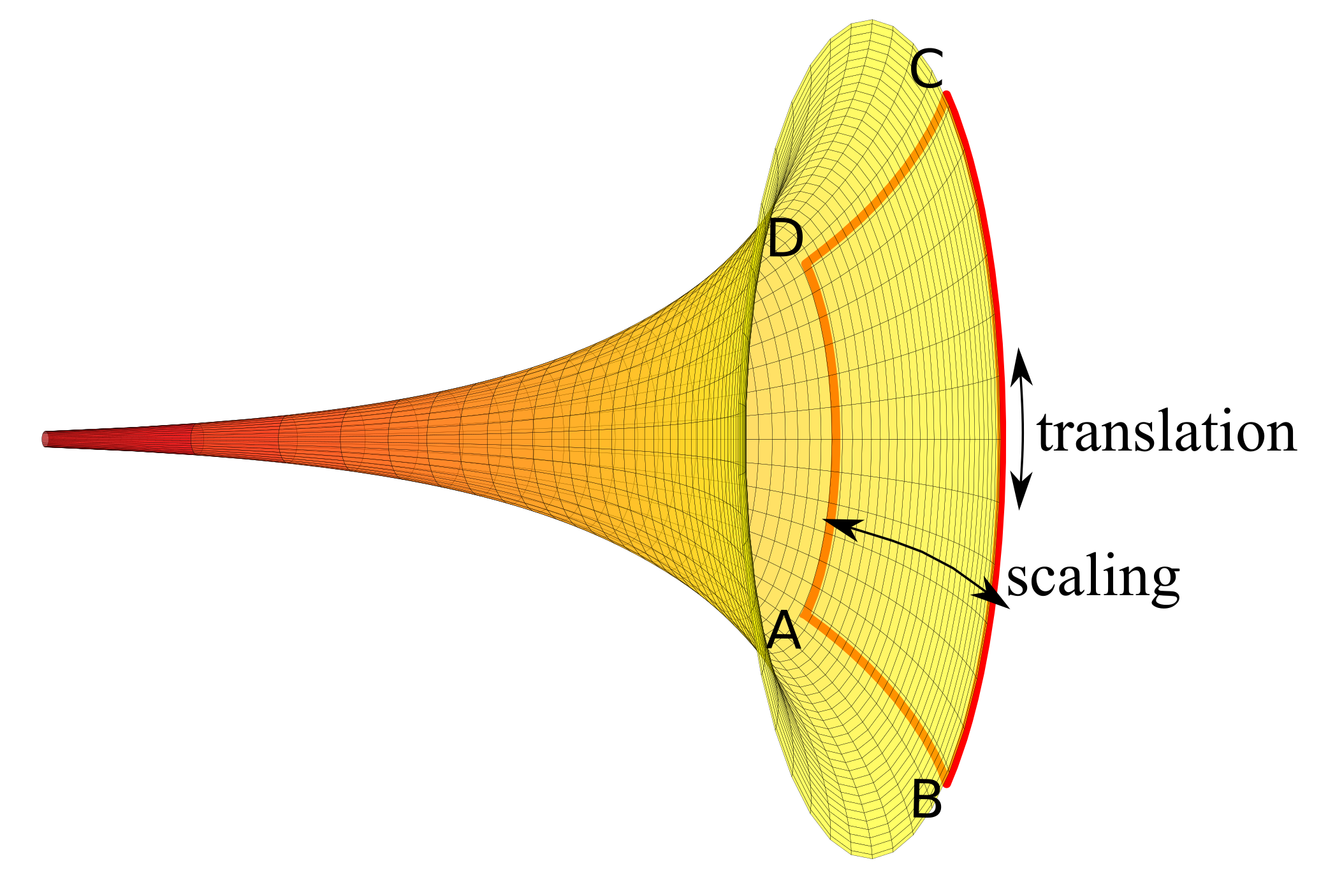}
    \put (05,65) {$(b)$}
    \end{overpic}
    \caption{(a) The parametric space of all possible states described by the parameters $a$ and $\Delta$ with marked red path of states described in the conditional wavefunction $\phi(x_2;x_1)$. (b) The embedding surface $\sigma$ of the Fubini-Study metric for the Gaussian function with marked red traversed path of states described by $\phi(x_2;x_1)$.}
    \label{fig:emb}
  \end{figure}  
  
  For some given functions $\Delta(x_1)$ and $a(x_1)$ we obtain a specific corresponding geometric potential $v^{\rm G}$, but the metric \eqref{eq:ds2-combi} allows us to see the hidden geometric aspects encoded in $v^{\rm G}$ that relate to any translation and scaling combination. 
  For example, for the Gaussian function $\phi_0(x_2) = 1/\pi^{1/4} e^{-x_2^2/2}$ the metric is
  \begin{align}
    ds^2 = \frac{1}{2} \left[ (d \ln \Delta)^2 + \frac{1}{\Delta^2} d a^2 \right] \, .
  \end{align}
  The metric can be visualized as embedding of a surface $\sigma$ in three dimensional space (see Fig.\ref{fig:emb}). 
  The surface $\sigma$ is described by the map
  \begin{align}
    \sigma: (a,\Delta) &\to (\rho,\varphi,z) \nonumber \\
    \rho(a,\Delta) &= \frac{1}{\Delta} \\
    \varphi(a,\Delta) &= a \\
    z(a,\Delta) &= \sqrt{\frac{1}{\Delta^2}-1}- \arctanh \sqrt{\frac{1}{\Delta^2}-1}\, ,
  \end{align}    
  where the mapping is written in the cylindrical coordinates $\rho$, $\varphi$ and $z$.
  The  scaling  of  the  state  occurs  in  the  radial  direction  along  the surface, while the translation of the state happens in the angular direction along the surface. 
  For given functions $a(x_1)$ and $\Delta(x_1)$, we get a curve $\gamma(x_1) = \sigma(a(x_1),\Delta(x_1)) : \mathbb{R} \rightarrow \mathbb{R}^3$ on the surface $\sigma$ and the distance measured along the curve corresponds to the traversed Fubini-Study distance (up to a factor $1/\sqrt{2}$),
  \begin{align}
    \int ds = \int  \sqrt{2v^{\rm G}(x_1)} \, dx_1 = \frac{1}{\sqrt{2}} \int  \left| \gamma'(x_1) \right| \, dx_1  \, .
  \end{align}

  For the concrete example (Fig. \ref{fig:comb}), the state is scaled down (A $\rightarrow$ B), translated up (B $\rightarrow$ C), scaled up (C $\rightarrow$ D) and translated down into original state (D $\rightarrow$ A). 
  Each transformation contributes to $v^{\rm G}$, but the greatest contribution is for the translation from state B to state C. 
  This is in agreement with the embedding picture Fig. \ref{fig:emb}, as the scaling down means moving out of the horn and afterwards the Fubini-Study metric is more sensitive to a translation. 
  Conversely, after scaling up (moving into the horn) a translation has a smaller contribution to the Fubini-Study metric.
  
  
  \subsection{Two state manifold}
  \label{sec:twostate}
  
  The previous transformations illustrate how the translation and the scaling of a state manifest in the Fubini-Study geometry. 
  Now, we study this geometry when the conditional wavefunction $\phi(x_2;x_1)$ along $x_1$ is fully described by a linear combination of two quantum states along $x_2$. 
  In such a case, the wavefunction can be written as
  \begin{align}
  \phi(x_2;x_1) = c_0(x_1) \phi_0 (x_2) + c_1(x_1) \phi_1 (x_2) \, , ~~~~ |c_0|^2 + |c_1|^2 = 1 \, ,
  \end{align}
  where $c_0$, $c_1$ are two parametric functions and $\phi_0$, $\phi_1$ are the wavefunctions of the two quantum states.
  
\begin{figure}[ht!]
    \begin{center}
      \includegraphics[scale=1.0]{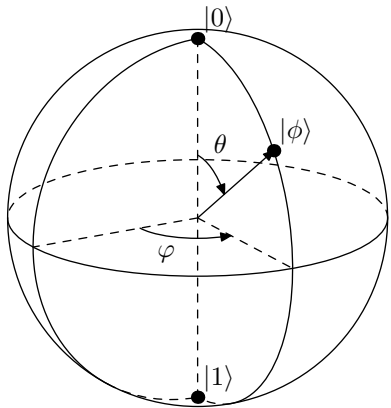}
    \end{center}
    \caption{Bloch sphere with coordinates $\theta$ and $\varphi$.} \label{fig:bloch01}
  \end{figure}

  This system has a well-known representation called the Bloch sphere (see Fig.\ \ref{fig:bloch01}) with a set of 2 natural coordinates
  \begin{align}
  \phi(x_2;x_1) = \cos\left[\theta(x_1)/2\right] \phi_0 (x_2) + e^{i \varphi(x_1)} \sin\left[\theta(x_1)/2\right] \phi_1 (x_2) \, ,
  \end{align}
  where $\theta \in [0,\pi ]$ and $\varphi \in [0, 2\pi )$ are two spherical angles.
  As in the previous section, the Fubini-Study metric $g_{11}$ can be seen as part of a two-dimensional Fubini-Study metric depending on the parameters $\theta$ and $\varphi$,
  \begin{align}
  	g_{11} (x_1) =
	\begin{pmatrix}
	\frac{d \theta}{d x_1 } & \frac{d \varphi}{d x_1 }
	\end{pmatrix}
	\begin{pmatrix}
	g_{\theta  \theta} & g_{\theta  \varphi} \\
    g_{\varphi \theta} & g_{\varphi \varphi}
    \end{pmatrix}
	\begin{pmatrix}
	\frac{d \theta }{d x_1 } \\
	\frac{d \varphi}{d x_1 }
	\end{pmatrix} \, ,
  \end{align}
  with
    \begin{align}
    g_{\theta  \theta} = \frac{1}{4} \, , ~~ g_{\theta  \varphi} = g_{\varphi \theta} = 0 \, , ~~ g_{\varphi \varphi} =  \frac{1}{4} \sin^2 \theta \, .
  \end{align}
  The Fubini-Study metric is then
  \begin{align}
  	d s^2 = \frac{1}{4} \left( d \theta^2 + \sin^2 \theta \, d \varphi^2 \right) \, ,
  \end{align}
  which corresponds to the natural metric on a sphere with radius $1/2$.
  
  In the case that this model represents an electronic system, there are symmetry conditions that have to be satisfied. 
  We now explain what happens in the case of two electrons.
  
  If we have a two-electron system restricted to two states, the symmetry-adapted electronic wavefunctions are of the form
  \begin{align}
   \psi^{\pm } (x_1,x_2) = \frac{1}{\sqrt{2}} \left( \phi_0 (x_1) \phi_1 (x_2) \pm \phi_1 (x_1) \phi_0 (x_2) \right) \, .
  \end{align}
  This form leads to the conditional wavefunction 
  \begin{align}
     \phi^{\pm } (x_2;x_1) = \frac{\phi_0 (x_1) \phi_1 (x_2) \pm \phi_1 (x_1) \phi_0 (x_2)}{\sqrt{  |\phi_0 (x_1)|^2+ |\phi_1 (x_1)|^2 }} \, , \label{eq:phipm}
  \end{align}    
  which gives us a direct mapping between the position $x_1$ and the coefficients $c_0$ and $c_1$, or the mapping between the position $x_1$ and the spherical angles $\theta$ and $\varphi$.
  
\begin{figure}[ht!]
    \begin{center}
      \includegraphics[scale=1.0]{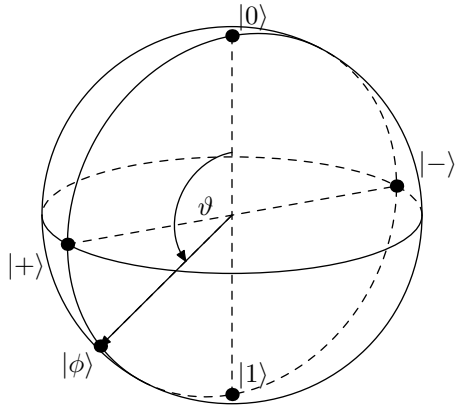}
    \end{center}
    \caption{Bloch sphere with coordinate $\vartheta$ and marked states $\ket{0}$ and $\ket{1}$. For later analysis, it is useful to mark also states $\ket{\pm} = \frac{1}{\sqrt{2}} \left(\ket{0} \pm \ket{1} \right)$.} \label{fig:bloch02}
  \end{figure}

  Often, the wavefunctions $\psi^\pm$ are real-valued functions. 
  In that case, the conditional wavefunction is also restricted to real values (the angle $\varphi$ is $0$ or $\pi$). This restriction corresponds to the great circle on the Bloch sphere going through the poles (see Fig.\ \ref{fig:bloch02}). 
  It is therefore suitable to change the coordinate system using one the central angle $\vartheta \in  [0,2\pi)$, such that
  \begin{align}
  \phi(x_2;x_1) = \cos\left[\vartheta(x_1)/2\right] \phi_0 (x_2) + \sin\left[\vartheta(x_1)/2\right] \phi_1 (x_2) \, . \label{eq:phicossin}
  \end{align}
  The Fubini-Study metric in this coordinate is simply
  \begin{align}
  d s^2 = \frac{1}{4} d \vartheta^2 \, . 
  \end{align}
  
  Equations \eqref{eq:phipm} and \eqref{eq:phicossin} provide an explicit mapping between $x_1$ and the central angle $\vartheta$,
  \begin{align}
    \cos \vartheta (x_1) = \frac{\phi_0(x_1)^2-\phi_1(x_1)^2}{\phi_0(x_1)^2+\phi_1(x_1)^2} \, , ~~~~~ 	\sin \vartheta (x_1) =  \frac{\pm 2 \phi_0(x_1) \phi_1(x_1)}{\phi_0(x_1)^2+\phi_1(x_1)^2} \, . \label{eq:maptheta}
  \end{align}
  Like in previous models, the traversed Fubini-Study distance $s$ can be obtained from the integral of $\sqrt{2 v^{\rm G}}$ as
  \begin{align}
    \int d s = \int^{x_1} \sqrt{2 v^{\rm G} (x_1')} d x_1' = \frac{1}{2} \int^{x_1} 
    \left|\frac{d \vartheta}{d x_1} \right| d x_1' = \frac{1}{2} \Theta (x_1) \, ,
  \end{align}
  where $\Theta (x_1)$ measures the traversed central angle $\vartheta$ on the great circle. How this function looks like depends solely on the states $\phi_1$ and $\phi_2$ and it is given by the mapping in equation \eqref{eq:maptheta}.

  \section{Diatomic model systems} 
  \label{sec:dia}
  
  The previous discussion helped us to understand the relationship between the structure of the conditional wavefunction of the electrons in the environment and the geometric potential. 
  In the following two sections, we study the behavior of $v^{\rm G}$ for one-dimensional models of a diatomic molecule with two electrons in the energetically lowest spatially antisymmetric state.
  This state corresponds to the lowest triplet state if spin was included.
  We select this state because for the spatially symmetric (singlet) ground-state there is no difference between the EEF and DFT, as only one DFT orbital is occupied.
  
  All model systems described in the following were solved numerically with the program package QMstunfti \cite{qmstunfti} that is based on the sparse-matrix functionality of Scipy \cite{scipy}, which in turn partially uses the ARPACK library \cite{arpack}.
  The exact KS potentials were obtained from the inversion procedure used in \cite{hodgson2017}, see also \cite{leeuwen1994}.
  We choose the gauge of zero vector potential, which is always possible for one-dimensional finite systems.
  
  \subsection{Model study of a one-dimensional homonuclear diatomic molecule}
  
  The Hamiltonian for a diatomic molecule in one dimension with clamped nuclei and two electrons is 
  \begin{align}
    H_{\rm m}^{(Z)}(R) = \sum_{j=1}^2 \left(-\frac{\partial_j^2}{2} + v_{\rm en}(x_j;R,Z) \right) + v_{\rm ee}(x_1,x_2) + v_{\rm nn}(R,Z).
    \label{eq:h_mol}
  \end{align}
  Here, we model all Coulomb interactions with soft-Coulomb potentials \cite{bauer1997}.
  Those potentials are often used in one-dimensional models to describe the effect that the Coulomb singularities can be avoided in three dimensions, but not in one dimension.
  The electron-nuclear interaction is therefore given by
  \begin{align}
    v_{\rm en}(x;R,Z) = -\frac{Z}{\sqrt{ (x+R/2)^2 + c_{\rm en} }} -\frac{1}{\sqrt{ (x-R/2)^2 + c_{\rm en} }},
  \end{align}
  corresponding to two nuclei with charges $Z$ and $+1$ located at $-R/2$ and $R/2$, respectively.
  The electron interaction is 
  \begin{align}
    v_{\rm ee}(x_1,x_2) = \frac{1}{\sqrt{ (x_1-x_2)^2 + c_{\rm ee} }}
    \label{eq:vee}
  \end{align}
  and the nuclear interaction is given by
  \begin{align}
    v_{\rm nn}(R,Z) = \frac{Z}{\sqrt{R^2 + c_{\rm nn}}}.
  \end{align}
  The external potential for this model is 
  \begin{align}
    v^{\rm ext}(x) &= v_{\rm en}(x;R,Z) + v_{\rm nn}(R,Z).
  \end{align}
  The parameters are \unit[$c_{\rm en} = c_{\rm ee} = 0.5$]{$a_0^2$} and \unit[$c_{\rm nn} = 0.1$]{$a_0^2$}.
  We first consider the homonuclear (symmetric) case with $Z=+1$.
  
  \begin{figure}[!htbp]
    \centering
    \includegraphics[width=.99\linewidth]{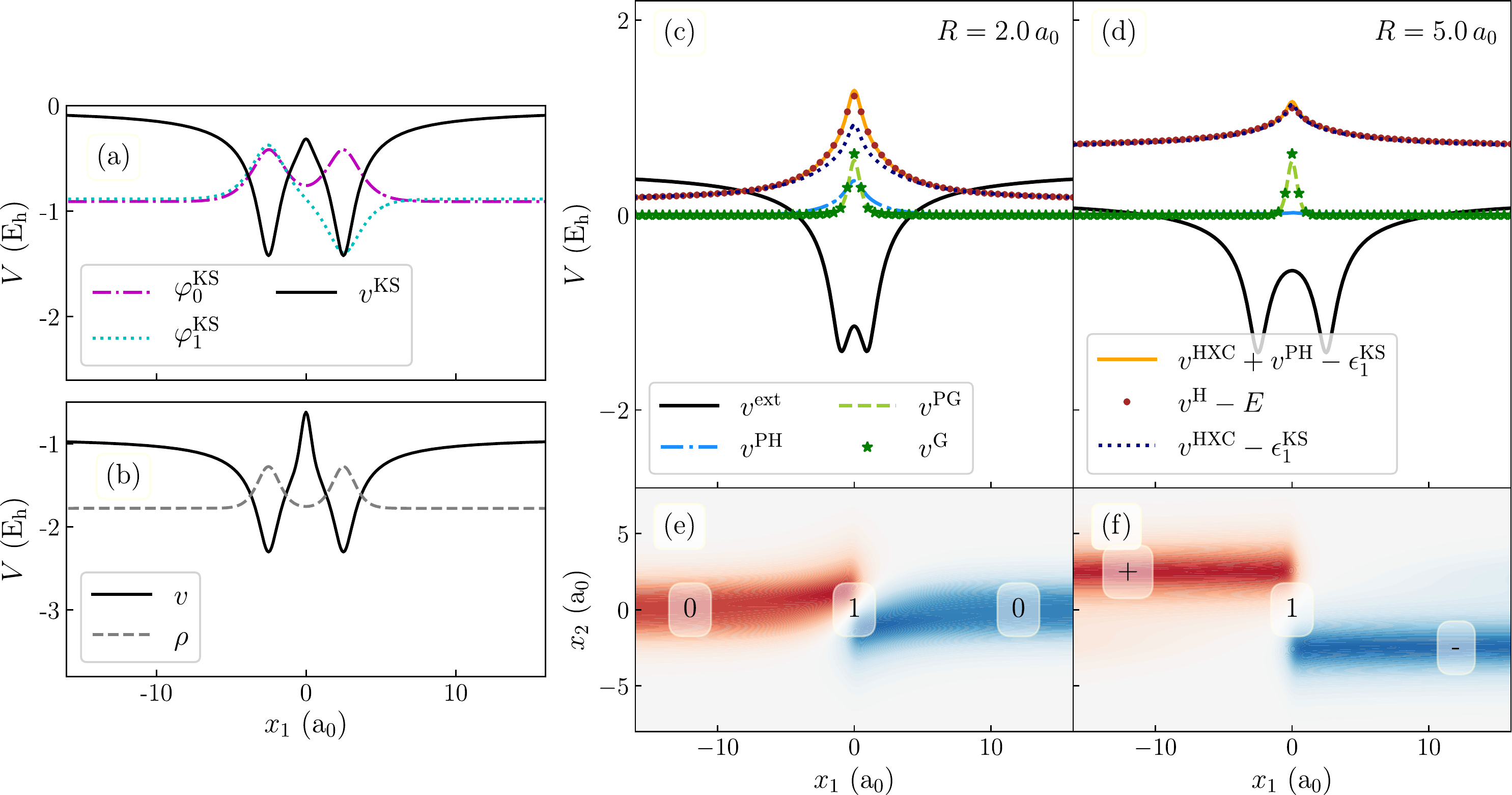}
    \caption{
    (a) KS potential $v^{\rm KS}$ and occupied KS orbitals $\varphi_j^{\rm KS}$ (shifted to their corresponding eigenvalues $\varepsilon_j^{\rm KS}$) as well as (b) EEF potential $v$ and one-electron density $\rho$ (shifted to its energy) for the energetically lowest antisymmetric electronic state of the homonuclear two-electron diatomic molecule with internuclear distance \unit[$R=5$]{$a_0$}.
    $v^{\rm KS}$ is shifted such that it is zero for large $|x_1|$, while the limit $|x_1| \rightarrow \infty$ of $v$ is the ground-state energy of the ionized system.
    (c), (d):
    Contributions to $v$ in the EEF and to DFT for two different internuclear distances $R$ of the homonuclear two-electron diatomic molecule, for its energetically lowest antisymmetric electronic state. 
    (e), (f): 
    Conditional wavefunctions $\phi(x_2;x_1)$ corresponding to the panels above, shown as contour plots (color indicates the sign).
    The states discussed in Sec.\ \ref{sec:twostatemodel} in the context of the geometric potential $v^{\rm G}$ are marked as ``0'', ``1'', ``+'', and ``--''.
    }
    \label{fig:pic_geo_sym}
  \end{figure}
  
  Fig.\ \ref{fig:pic_geo_sym}a shows the exact KS potential of the model together with the two lowest KS orbitals for an internuclear distance of \unit[$R = 5$]{$a_0$}, and Fig.\ \ref{fig:pic_geo_sym}b shows the EEF potential together with the one-electron density.
  The KS-orbitals are almost degenerate and look like typical tunneling states of a double well potential, with the wavefunction of the energetically lower state having the same sign in both wells, while the wavefunction of the energetically higher state switches sign at $x_1=0$.
  
  The components of the EEF potential $v$ and the KS-DFT potentials are depicted in Fig.\ \ref{fig:pic_geo_sym}c and \ref{fig:pic_geo_sym}d for the internuclear distances \unit[$R = 2$]{$a_0$} and \unit[$R = 5$]{$a_0$}, respectively.
  For the chosen parameters, the model is well-described by Hartree-Fock theory and the KS wavefunction is very close to the interacting wavefunction.
  Thus, the parts of $v$ based on the KS quantities and in the EEF are similar,
  \begin{subequations}
    \begin{align}
      v^{\rm PG} &\approx v^{\rm G} \\
      v^{\rm PH} + v^{\rm HXC} &\approx v^{\rm H}
    \end{align}
    \label{eq:pksimfs}
  \end{subequations} 
  with the second relation holding up to a constant.
  Some small differences of up to \unit[0.08]{$E_h$} still exists, with  $v^{\rm PG}$ being slightly smaller than $v^{\rm G}$ and $v^{\rm PH} + v^{\rm HXC} - \varepsilon_1^{\rm KS}$ being slightly larger than $v^{\rm H} - E$, because Hartree-Fock theory does not describe the system exactly.
  In appendix \ref{sec:ecorr} we provide an example where the differences between the potentials is larger.
  
  Let us now turn to the features of the potentials an their origin in terms of the state $\phi$ of the electron in the environment.
  The wavefunctions $\phi(x_2;x_1)$ of the environment for the two internuclear distances are shown in Fig.\ \ref{fig:pic_geo_sym}e and \ref{fig:pic_geo_sym}f.
  As $\phi(x_2;x_1)$ is the wavefunction of one electron at $x_2$ given there is another one at $x_1$, we have that for a large internuclear distance $R$ the electron at $x_2$ is either located at one nucleus or at the other.
  For \unit[$R=5$]{$a_0$}, given one electron of the two-electron system is found at, say, \unit[$x_1<-2$]{$a_0$}, it is likely to ``originate'' from the nucleus at \unit[$x_1=-2.5$]{$a_0$}.
  The second electron is thus most likely found at the other nucleus centered around \unit[$x_2=+2.5$]{$a_0$} and $\phi(x_2;x_1)$ corresponds approximately to the ground state of an electron at that nucleus.
  
  This behavior is reflected in the potentials.
  All parts of $v$ except the external potential $v^{\rm ext}$ are repulsive bell-shaped potentials centered around $x_1 = 0$.
  For \unit[$R = 2$]{$a_0$}, the two electrons of the system are relatively close to each other.
  When the interatomic distance $R$ is increased, the system approaches the limit of two separated one-electron atoms.
  For \unit[$R = 5$]{$a_0$}, $v^{\rm PH}$ is already almost zero, indicating that the energy of the environment in the KS system does not depend on the location of the electron at $x_1$.
  This can be interpreted such that when the electron at $x_1$ is at one nucleus, the electron in the environment is always at the other nucleus and is essentially undisturbed.
  However, in the interacting system the two electrons still influence each other, as is visible in $v^{\rm H}$:
  It becomes smaller for larger $R$ (and eventually becomes a constant shift for $R \rightarrow \infty$), but it is showing that the position of the electron at $x_1$ matters for the energy of the environment, because $v^{\rm H}$ is still a repulsive bell-shaped curve.
  In the KS system, in contrast, this effect is fully contained in $v^{\rm HXC}$.
  
  Turning to our main interest, the geometric potential $v^{\rm G}$, we observe what is expected from the discussion in Section \ref{sec:qgt}:
  The potential $v^{\rm G}$ is a measure of how strong the wavefunction $\phi(x_2;x_1)$  of the one-electron environment changes if the other electron is moved along $x_1$.
  There is a large change in the conditional wavefunction only at $x_1 \approx 0$ that is reflected in $v^{\rm G}$ as a peak in this region (see also \cite{gritsenko1996b,giarrusso2018} for such a peak in similar models).
  If the internuclear distance is increased, the system becomes more and more that of two separated atoms and the change of the conditional wavefunction at $x_1 \approx 0$ becomes sharper.
  With increasing internuclear distance the peak of the geometric potential $v^{\rm G}$ (or $v^{\rm PG}$) then becomes somewhat more localized at $x_1=0$ and also slightly higher, although this is hardly visible in the figure.
  This peak indicates that when the electron at $x_1$ is moved along $x_1$, it is around $x_1 \approx 0$ that the electron in the environment jumps from one nucleus to the other.
  We emphasize that the shape of $v^{\rm G}$ has nothing to do with the sign change of $\phi(x_2;x_1)$ along $x_1$, as can been seen from the definition \eqref{eq:eef_vfs} of $v^{\rm G}$.
  It is for example also present in the symmetric ground state of $H_{\rm m}^{(1)}$ for which no sign change in $\phi(x_2;x_1)$ happens but $\phi(x_2;x_1)$ is otherwise similar (cf.\ \cite{buijse1989,benitez2016})
  
  \subsection{Model study of a one-dimensional heteronuclear diatomic molecule}
  
  We now investigate the effect of having a heteronuclear diatomic molecule instead of a homonuclear one.
  For this purpose, we use $Z=2$ in the Hamiltonian \eqref{eq:h_mol}, such that there is one nucleus at $+R/2$ with charge $+1$ and one nucleus at $-R/2$ with charge $+2$, and we again consider the lowest antisymmetric state.
  We used the same model already in \cite{complet} to study the appearance of charge-transfer steps in the EEF and in DFT, and we found the origin of the steps in the energy $v^{\rm H}$ of the environment.
  Here, we use it again to study the geometric potential $v^{\rm G}$.
  
  \begin{figure}[!htbp]
    \centering
    \includegraphics[width=.99\linewidth]{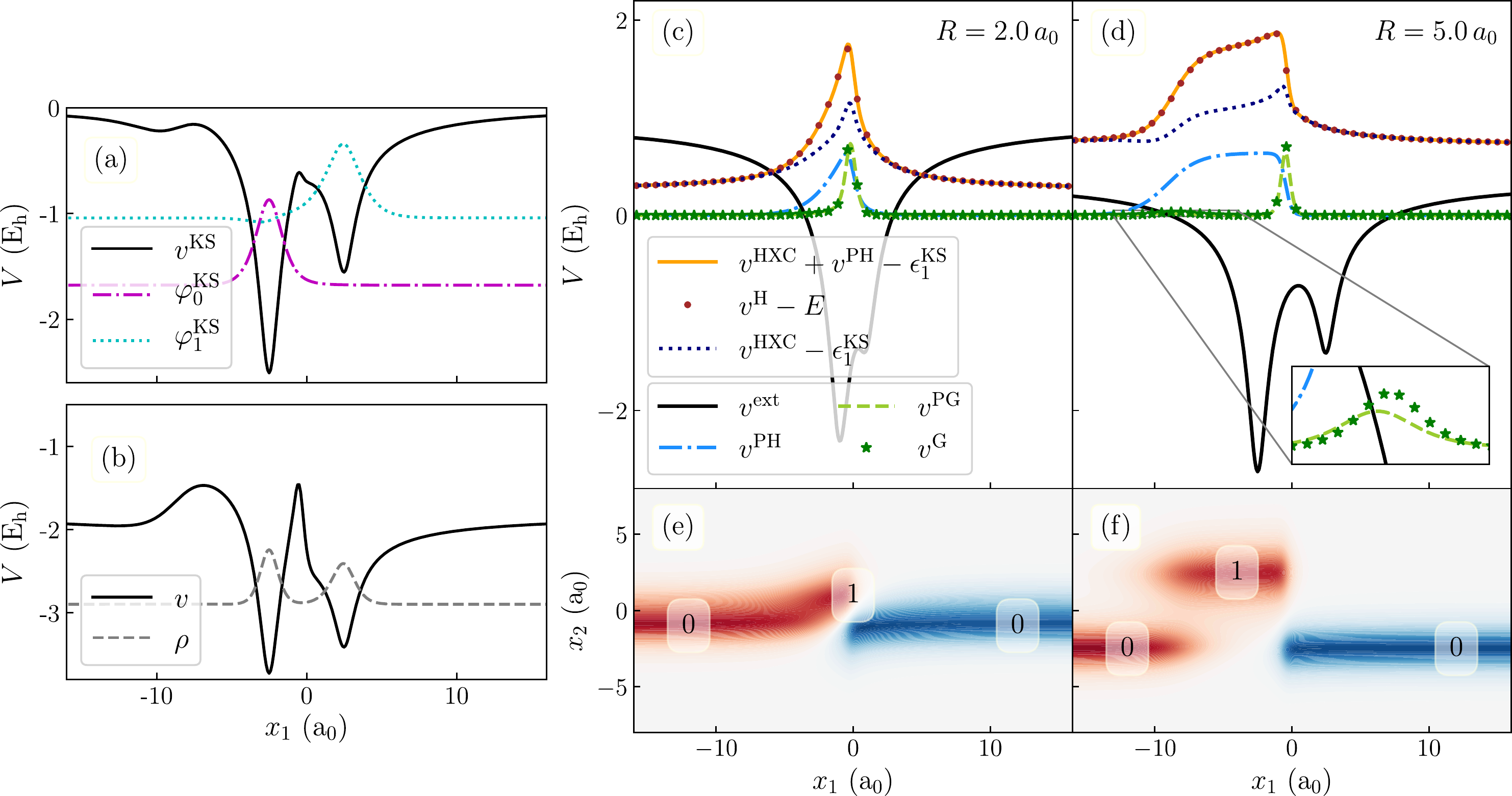}
    \caption{
    (a) KS potential $v^{\rm KS}$ and occupied KS orbitals $\varphi_j^{\rm KS}$ (shifted to their corresponding eigenvalues $\varepsilon_j^{\rm KS}$) as well as (b) EEF potential $v$ and one-electron density $\rho$ (shifted to its energy) for the energetically lowest antisymmetric electronic state of the heteronuclear two-electron diatomic molecule with internuclear distance \unit[$R=5$]{$a_0$}.
    $v^{\rm KS}$ is shifted such that it is zero for large $|x_1|$, while the limit of $|x_1| \rightarrow \infty$ of $v$ is the ground-state energy of the ionized system.
    (c), (d): 
    Contributions to $v$ in the EEF and based on KS quantities for two different internuclear distances $R$ of the heteronuclear two-electron diatomic molecule, for the energetically lowest antisymmetric electronic state. 
    In (d), an inset shows the details of $v^{\rm G}$ for $x_1 \in [-13,-4] \, a_0$.
    (e), (f): 
    Conditional wavefunctions $\phi(x_2;x_1)$ corresponding to the panels above, shown as contour plots (color indicates the sign).
    The states ``0'' and ``1'' discussed in Sec.\ \ref{sec:twostatemodel} in the context of the geometric potential $v^{\rm G}$ are marked.
    }
    \label{fig:pic_geo_asy}
  \end{figure}
  
  Fig.\ \ref{fig:pic_geo_asy}a shows the KS potential together with the relevant KS orbitals and Fig.\ \ref{fig:pic_geo_asy}b shows the EEF potential together with the one-electron density for an internuclear distance of \unit[$R = 5$]{$a_0$}.
  The lowest KS orbital $\vp_0^{\rm KS}$ is localized around \unit[$x_1 = -2.5$]{$a_0$} (where the nucleus with charge +2 is located) while the highest occupied KS orbital $\vp_1^{\rm KS}$ is localized around \unit[$x_1 = +2.5$]{$a_0$} (where the nucleus with charge +1 is located).
  
  The contributions to the EEF and KS potentials are depicted in panels Fig.\ \ref{fig:pic_geo_asy}c and \ref{fig:pic_geo_asy}d for the internuclear distances \unit[$R = 2$]{$a_0$} and \unit[$R = 5$]{$a_0$}, respectively.
  Also this model is well-described by Hartree-Fock theory, hence Relations \eqref{eq:pksimfs} hold (see Appendix \ref{sec:ecorr} for parameters of the model where these relations do not hold).
  For larger distances, $v^{\rm H}$ and $v^{\rm HXC} + v^{\rm PH}$ form a step, because the state of the environment changes along $x_1$.
  For a detailed discussion of this charge-transfer step and its interpretation, see \cite{complet}.
  
  Here, we focus on the  geometric potential $v^{\rm G}$.
  It looks similar to that of the homonuclear diatomic, with a bell-shaped maximum centered at $x_1 \approx 0$ that indicates a qualitative change of the conditional wavefunction $\phi(x_2;x_1)$, depicted in Fig.\ \ref{fig:pic_geo_asy}e and \ref{fig:pic_geo_asy}f, respectively.
  However, for larger internuclear distances there is a second significant change of $\phi(x_2;x_1)$ along $x_1$.
  For \unit[$R = 5$]{$a_0$} this second change is at \unit[$x_1 \approx -8$]{$a_0$}.
  If an electron is found somewhere in $-8 \, a_0 < x_1 < 0$, we see from $\phi(x_2;x_1)$ that the second electron is most probably found around \unit[$x_2 = 2.5$]{$a_0$}, corresponding to the location of the nucleus with charge $+1$.
  In contrast, if an electron is found at \unit[$x_1 < -8$]{$a_0$}, the second electron is found around \unit[$x_2 = -2.5$]{$a_0$}, corresponding to the nucleus with charge $+2$.
  The larger the internuclear distance $R$ becomes, the more does this transition move to smaller values of $x_1$ (i.e., to the left).
  Mechanistically, what happens at the location of the peak is a charge transfer, where the electron at $x_2$ switches from the energetically  higher potential well to the lower well depending on where the other electron at $x_1$ is located.  \cite{complet}
  Close inspection of the potentials for \unit[$R = 5$]{$a_0$} reveals that this charge transfer is visible in $v^{\rm G}$, albeit barely: 
  It leads to a second peak in $v^{\rm G}$, as shown in the inset of \ref{fig:pic_geo_asy}d.
  As the change of $\phi$ in that region happens over a rather large distance along $x_1$ compared with the change at $x_1 \approx 0$, it leads to a small but broad local increase in $v^{\rm G}$.
  This second peak in $v^{\rm G}$ has also been found in a similar model recently \cite{giarrusso2018}, where it was concluded that the peak is at the side of the more electronegative atom, in agreement with our interpretation.
  We note that although there is a large difference in the height and width of the peak, the two changes of the conditional wavefunction are rather similar.
  This leads to the integrals of $\sqrt{v^{\rm G}}$ over the corresponding regions to both be approximately equal, as explained in the next section.
  
  \subsection{Two-state analysis}
  \label{sec:twostatemodel}
  
  The two parts $v^{\rm H}$ and $v^{\rm G}$ of the EEF potential $v$ are functionals of the conditional wavefunction $\phi$.
  By looking at the conditional wavefunctions $\phi$ for the symmetric and asymmetric diatomic in Fig.\ \ref{fig:pic_geo_sym} and in Fig.\ \ref{fig:pic_geo_asy}, respectively, is seems that $\phi(x_2;x_1)$ for some value of $x_1$ can be described as a two-state problem.
  This view is also supported by the observation that all our models are well-described with the Hartree-Fock approximation and the Hartree-Fock and Kohn-Sham orbitals differ very little.
  Hence, the wavefunction of the presented two-electron models of the diatomic molecule can approximately be written as
  \begin{align}
    \psi(x_1,x_2) \approx \frac{1}{\sqrt{2}} \left( \vp_0^{\rm KS}(x_1) \vp_1^{\rm KS}(x_2) 
                                                  - \vp_1^{\rm KS}(x_1) \vp_0^{\rm KS}(x_2) \right)
                                                  \label{eq:hfapprox}
  \end{align}
  with the conditional wavefunction
  \begin{align}
    \phi(x_2;x_1) \approx \frac{ \vp_0^{\rm KS}(x_1) \vp_1^{\rm KS}(x_2) 
                                                  - \vp_1^{\rm KS}(x_1) \vp_0^{\rm KS}(x_2) }{\sqrt{|\vp_0^{\rm KS}(x_1)|^2 + |\vp_1^{\rm KS}(x_1)|^2}}
                                                  \label{eq:hfapproxc}
  \end{align}
  for the gauge $\chi(x_1) = \sqrt{\rho(x_1)}$. Thus, our models are very well described by the two-state model from Sec.\ \ref{sec:twostate} with states $\ket{0}$ and $\ket{1}$ in Figs.\ \ref{fig:bloch01} and \ref{fig:bloch02} chosen as the KS orbitals $\varphi^{\rm KS}_0$ and $\varphi^{\rm KS}_1$.
 
  Then, the map between the electron coordinate $x_1$ and the central angle $\vartheta$ (see Fig.\  \ref{fig:bloch02}) is
  \begin{equation}
  \cos \vartheta (x_1) = \frac{\varphi^{\rm KS}_0(x_1)^2 - \varphi^{\rm KS}_1(x_1)^2}{\varphi^{\rm KS}_0(x_1)^2 + \varphi^{\rm KS}_1(x_1)^2} \, , ~~~~ 
    \sin \vartheta (x_1) = \frac{- 2 \varphi^{\rm KS}_0(x_1) \varphi^{\rm KS}_1(x_1)}{\varphi^{\rm KS}_0(x_1)^2 + \varphi^{\rm KS}_1(x_1)^2} \, ,
    \label{eq:thetav2}
  \end{equation}
  and the geometric potential $v^{\rm G}$ is 
  \begin{equation}
    v^{\rm G} (x_1) = \frac{1}{8} \left(\pa \vartheta\right)^2 \, .
  \end{equation}
  For the considered KS orbitals, the resulting map $\vartheta(x_1)$ is a monotonic function. 
  Due to this monotonicity we get another relation for the central angle,
  \begin{equation}
    \vartheta(x_1) = \int_{-\infty}^{x_1} \sqrt{8 v^{\rm{G}} (x_1')} \mathop{d x_1'} \, .
    \label{eq:theta}
  \end{equation}
  
  \begin{figure}[!htbp]
    \centering
    \includegraphics[width=.99\linewidth]{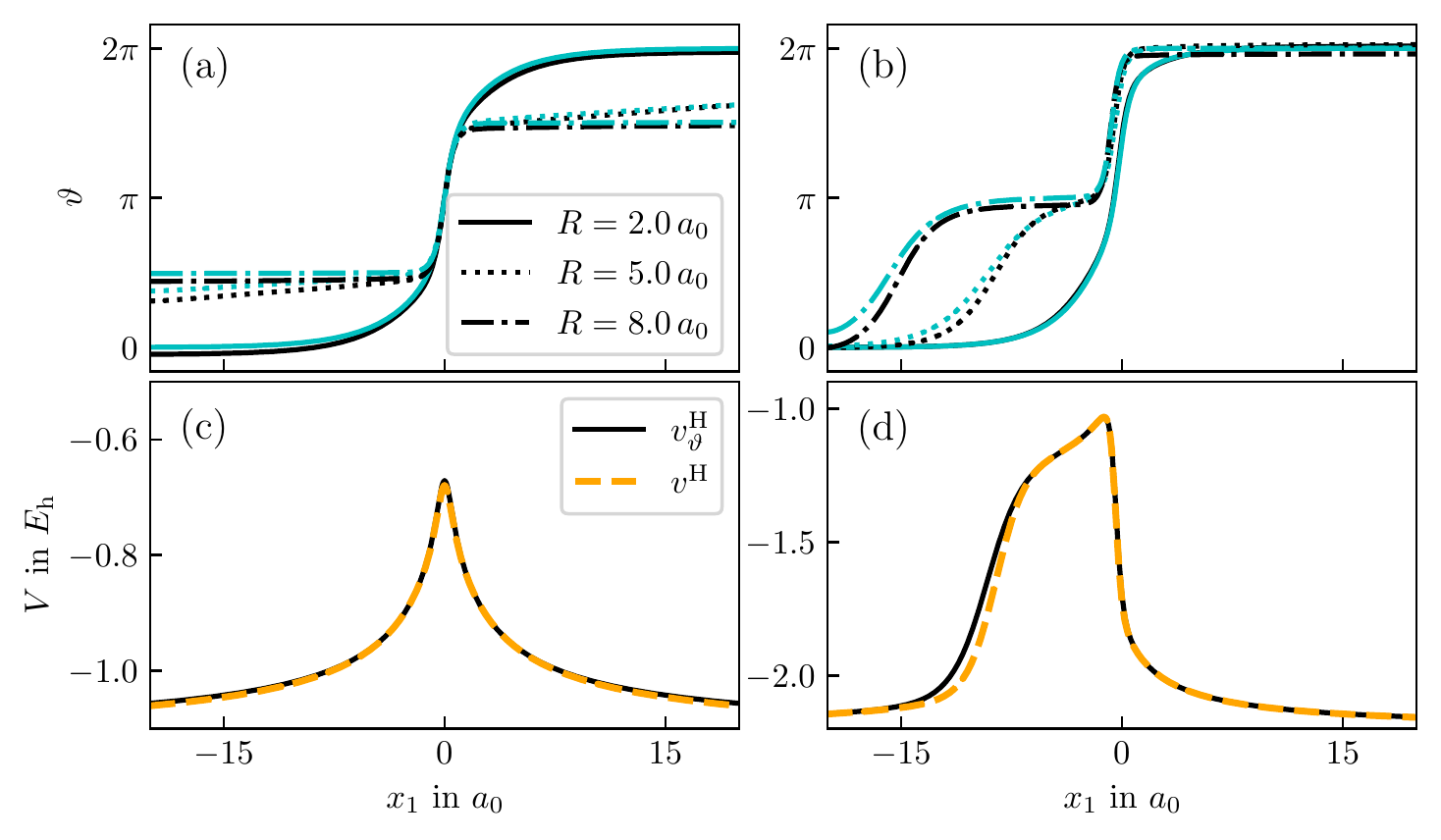}
    \caption{
    Top: Angle $\vartheta$ determined from \eqref{eq:theta} (black lines) and from \eqref{eq:thetav2} (cyan lines) for different values of the internuclear separation $R$ for (a) the homonuclear diatomic and (b) the heteronuclear diatomic.
    Bottom: Energy of the environmental electron $v^{\rm H}$ and $v_{\rm \vartheta}^{\rm H}$ determined by \eqref{eq:vHrec} with $\vartheta$ being determined by \eqref{eq:thetav2}, for \unit[$R=5.0$]{$a_0$}, for (c) the homonuclear diatomic and (d) the heteronuclear diatomic.}
    \label{fig:pic_state_angle}
  \end{figure}
  
  There are thus two ways to compute the parameter $\vartheta(x_1)$ that determines the conditional wavefunction $\phi(x_2; x_1)$ in the two-state model:
  One is based on the KS orbitals, \eqref{eq:thetav2}, and one is based on the geometric potential, \eqref{eq:theta}.
  The parameter $\vartheta$ obtained in both ways is shown in Fig.\ \ref{fig:pic_state_angle}a and Fig.\ \ref{fig:pic_state_angle}b for the homonuclear and heteronuclear diatomic, respectively.
  There is little difference between the two ways of obtaining $\vartheta$ for both the model of the homo- and heteronuclear diatomic, i.e., the two-state approximation is justified.
  
  The behavior of $\vartheta$ reflects the behavior of $\phi$ depicted in Figs.\ \ref{fig:pic_geo_sym} and \ref{fig:pic_geo_asy}.
  To interpret $\phi$, some regions in these figures were marked with ``$0$'', ``$1$'', ``$+$'' or ``$-$''.
  These represent states between which $\phi(x_2; x_1)$ changes along $x_1$.
  For the homonuclear diatomic with small internuclear distance, $\phi(x_2; x_1)$ for $|x_1| \rightarrow \infty$ is close to the energetically lowest KS orbital $\vp_0^{\rm KS}$ and resembles the wavefunction of the ionized system.
  At $x_1 = 0$, it follows from \eqref{eq:hfapproxc} and the fact that $\vp_1^{\rm KS}(0) = 0$ (since $\vp_1^{\rm KS}$ is even) that $\phi(x_2; x_1=0) \approx \vp_1^{\rm KS}(x_2)$.
  Hence, we have that $\phi(x_2; x_1)$ changes as 
  \begin{align}
    \phi(x_2; -\infty) \approx \vp_0^{\rm KS} 
      ~ \Rightarrow ~ \phi(x_2; 0) \approx \vp_1^{\rm KS}
      ~ \Rightarrow ~ \phi(x_2; +\infty) \approx \vp_0^{\rm KS}
      \label{eq:pathsym1}
  \end{align}
  (the sign of the state does not matter in the metric).
  This behavior is valid e.g.\ for an internuclear distance \unit[$R=2$]{$a_0$}, see Fig.\ \ref{fig:pic_geo_sym}e, where state ``0'' is approximately $\vp_0^{\rm KS}$ and ``1'' is approximately $\vp_1^{\rm KS}$.
  The path \eqref{eq:pathsym1} of $\phi$ from $x_1 = -\infty$ to $x_1 = +\infty$ corresponds to one full motion along the great circle, i.e., to a total change 
  \begin{align}
    |\vartheta(x_1=+\infty) - \vartheta(x_1=-\infty)| \approx 2 \pi
  \end{align}
  of the parameter $\vartheta$, as shown in Fig.\ \ref{fig:pic_state_angle}a for \unit[$R=2$]{$a_0$}.
  
  In contrast, for larger internuclear distances, $\phi(x_2; x_1)$ for $|x_1| \rightarrow \infty$ is either located on one or the other nucleus.
  Due to the symmetry of the problem, this corresponds approximately to the two states 
  \begin{align}
   \vp_{\pm}^{\rm KS} = \frac{1}{\sqrt{2}} \left( \vp_0^{\rm KS} \pm \vp_1^{\rm KS} \right)
  \end{align}
  which are on opposite sides of the Bloch sphere at the equator ($\ket{\pm}$ in Fig.\ \ref{fig:bloch02}), if $\vp_0^{\rm KS}$ and $\vp_1^{\rm KS}$ are the poles ($\ket{0}$ and $\ket{1}$). 
  Equation \eqref{eq:hfapproxc} is still valid and $\phi(x_2; x_1=0) \approx \vp_1^{\rm KS}(x_2)$, hence we have 
  \begin{align}
    \phi(x_2; x_1 \ll 0) \approx \vp_{+}^{\rm KS}
      ~ \Rightarrow ~ \phi(x_2; 0) \approx \vp_1^{\rm KS}
      ~ \Rightarrow ~ \phi(x_2; x_1 \gg 0) \approx \vp_{-}^{\rm KS}
      \label{eq:pathsym2}
  \end{align}
  which is visible in Fig.\ \ref{fig:pic_geo_sym}f, where state ``$+$'' is approximately $\vp_{+}^{\rm KS}$ and state ``$-$'' is approximately $\vp_{-}^{\rm KS}$.
  The path \eqref{eq:pathsym2} corresponds to half of a great circle on the Bloch sphere, i.e.,
  \begin{align}
    |\vartheta(x_1 \gg 0) - \vartheta(x_1 \ll 0)| \approx \pi,
  \end{align}
  as shown in Fig.\ \ref{fig:pic_state_angle}a for \unit[$R=5$]{$a_0$} and \unit[$8$]{$a_0$}.
  
  For the heteronuclear diatomic at small internuclear distances the situation is similar to that of the homonuclear diatomic, see Fig.\ \ref{fig:pic_geo_asy}e.
  In contrast, at larger internuclear distances, the KS orbitals are either localized on one or the other nucleus.
  From Fig.\ \ref{fig:pic_geo_asy}f we can see that there is a hopping of $\phi(x_2;x_1)$ from one side to the other and back along $x_1$, as discussed above, which is
  \begin{align}
    \phi(x_2; -\infty) \approx \vp_0^{\rm KS}
      ~ \Rightarrow ~ \phi(x_2; x_{\rm S}) \approx \vp_1^{\rm KS}
      ~ \Rightarrow ~ \phi(x_2; +\infty) \approx \vp_0^{\rm KS}
  \end{align}
  with $x_{\rm S}$ representing the corresponding region of $x_1$.
  Thus, 
  \begin{align}
    |\vartheta(x_1=+\infty) - \vartheta(x_1=-\infty)| \approx 2 \pi.
  \end{align}
  for any finite value of the internuclear distance $R$ but happens in two steps which each correspond to a change of $\pi$, with one smooth change somewhere at $x_1<0$ and a sharp change at $x_1\approx0$, as can be seen in Fig.\ \ref{fig:pic_state_angle}b.
  The smooth change moves more and more towards negative $x_1$ when the internuclear distance is increased.
  
  The fact that the asymptotic behavior of the KS orbitals is known can be used to construct $v^{\rm G}$ analytically in the region where the dominant orbital changes, at least if this asymptotic behavior is attained.
  The leading term in the asymptotic KS orbitals for $\lim\limits_{|x_1| \rightarrow \infty} v^{\rm KS}(x_1) = 0$ is
  \begin{equation}
    \varphi^\mathrm{KS}_i (x_1) \propto e^{-\sqrt{-2 \varepsilon^\mathrm{KS}_i} |x_1|}.
    \label{eq:ksorbasy}
  \end{equation}
  The dominant orbital changes, say, around $x_{\rm c}$, and we write the KS orbitals as
  \begin{equation}
    \varphi_i (x_1) = C e^{-A_i (x_1-x_{\rm c})} \, ,
  \end{equation}
  where $A_i$ is $\pm \sqrt{-2 \varepsilon^\mathrm{KS}_i}$ with the sign depending on the direction of the decay. 
  Using the explicit form of the KS orbitals in \eqref{eq:thetav2} leads to
  \begin{equation}
    \cos \vartheta = \frac{1 - e^{-2 \Delta (x_1-x_{\rm c})}}{1 + e^{-2 \Delta (x_1-x_{\rm c})}} \, , ~~~~~~~~ \sin \vartheta = \frac{2 e^{-\Delta (x_1-x_{\rm c})}}{1 + e^{-2 \Delta (x_1-x_{\rm c})}} \, ,
  \end{equation}
  where $\Delta = A_i - A_j$ with $i$ and $j$ being the indices of the two involved KS orbitals.
  We thus find that the geometric potential is a bell-shaped function with the width $1/\Delta$ and the height $\Delta^2/8$,
  \begin{equation}
    v^{\rm G} (x_1) = \frac{1}{8} \left(\pa \vartheta \right)^2 =   \frac{1}{8} \left( \frac{1}{ \sin \vartheta} \pa \cos (\vartheta) \right)^2 = \frac{1}{8} \Delta^2 \sech^2(\Delta (x_1-x_{\rm c})) \, .
    \label{eq:vga}
  \end{equation}
  The integral
  \begin{equation}
    \int \sqrt{8 v^{\rm G} (x_1)} \, dx_1 = \gd (\Delta (x_1-x_{\rm c})) \, ,
  \end{equation}
  is the Gudermannian function $\gd$ \cite{HORADAM196844} and yields
  \begin{equation}
    \int_{-\infty}^{+\infty} \sqrt{8 v^{\rm G} (x_1)} \, dx_1 = \pi.
    \label{eq:pipi}
  \end{equation}
   Relation \eqref{eq:pipi} means that the integral of $\sqrt{8 v^{\rm G}}$ over the region of a change of the KS orbital which dominates the density is equal to $\pi$.
   The analytic form \eqref{eq:vga} of $v^{\rm G}$ is not restricted to our models but is general, provided the KS orbitals can be approximately described by \eqref{eq:ksorbasy} and provided there is only a switch from one dominant orbital to another, which is typically the case.
  
  If the internuclear distance is large enough, e.g.\ for \unit[$R=5$]{$a_0$} and for \unit[$R=8$]{$a_0$}, there is one such region for the homonuclear diatomic model and two such regions for the heteronuclear model.
  For the heteronuclear diatomic, the shape of the geometric potential of the region outside the internuclear region is a broad and shallow bell with parameter $\Delta = \sqrt{-2 \varepsilon^\mathrm{KS}_0} - \sqrt{-2 \varepsilon^\mathrm{KS}_1}$, see the inset in Fig.\ \ref{fig:pic_geo_asy}d.
  For both the homo- and heteronuclear diatomic, the geometric potential inside the internuclear region is a narrow and high bell, see Fig.\ \ref{fig:pic_geo_asy}c and \ref{fig:pic_geo_asy}d, respectively.
  If the internuclear distance $R$ is large enough, the corresponding parameter is $\Delta = \sqrt{-2 \varepsilon^\mathrm{KS}_0} + \sqrt{-2 \varepsilon^\mathrm{KS}_1}$ (for the homonuclear case also $\varepsilon^\mathrm{KS}_0 \approx \varepsilon^\mathrm{KS}_1$).
  The considered values of $R$ (up to ca.\ \unit[$R=10$]{$a_0$}, then the electron density in the internuclear region becomes too small and numerical artefacts appear) are still too small for this relation to hold, but \eqref{eq:vga} is valid, albeit with different $\Delta$.
  
  Finally, we note that we can use the two-state assumption to construct $v^{\rm H}$ from $v^{\rm G}$ (and vice versa).
  For this purpose, the expression \eqref{eq:vga} can be used with \eqref{eq:theta} to determine $\vartheta$
  Then, the conditional wavefunction $\phi$ can be obtained from \eqref{eq:phicossin} after identifying $\phi_j = \varphi_j^{\rm KS}$, and the energy of the environment $v^{\rm H}$ can be obtained from its definition \eqref{eq:eef_vh} as
  \begin{equation}
  v^{\rm H} (x_1) \approx v_{\vartheta}^{\rm H} (x_1) 
  = \frac{h_{00} + h_{11}}{2} + \frac{h_{00} - h_{11}}{2} \cos \vartheta (x_1) +  h_{01} \sin \vartheta (x_1).
  \label{eq:vHrec}
  \end{equation}
  Here, $h_{ij}$ correspond to matrix elements of the operator evaluated for $v^{\rm H}$ in the chosen basis $\ket{0}$ and $\ket{1}$.
  In the basis of the KS orbitals and for the model potential, the matrix elements are
  \begin{align}
    h_{ij} = \Braket{\vp_i^{\rm KS}(2) | -\frac{\pb^2}{2} + v_{\rm en}(2) + v_{\rm ee}(1,2) | \vp_j^{\rm KS}(2)}_2.
  \end{align}
  The only $x_1$-dependence of $h_{ij}$ is due to the electron-electron interaction $v_{\rm ee}$.
  Fig.\ \ref{fig:pic_state_angle}c and \ref{fig:pic_state_angle}d illustrate that the reconstruction is very accurate for \unit[$R=5$]{$a_0$}:
  From $v^{\rm G}$ the angle $\vartheta$ can be determined via \eqref{eq:theta}, and this angle can be used in \eqref{eq:vHrec} to obtain $v_{\vartheta}^{\rm H}$, which is very close to the true energy $v^{\rm H}$ of the environment.
  This reconstruction works equally well for the other considered internuclear distances.
  
  \section{Summary}
  \label{sec:sum}
  
  The EEF is an exact one-electron theory of the many electron problem.
  It provides a clear and intuitive picture of one electron in the environment of other electrons.
  The wavefunction of those other electrons $\phi(2, \dots, N;1)$ provides the scalar potential $v$ and the vector potential that appear in the one-electron Schr\"odinger equation, where $v$ is the sum of $v^{\rm H}$, the energy of the environment in the presence of an additional electron, and $v^{\rm G}$, the geometric potential.
  
  Generally, in the EEF the interacting many-electron wavefunction is considered.
  If the KS wavefunction is used instead, the same one-electron potential $v$ is obtained but the contributions to $v$ are different due to the different way of how the electronic structure is described.
  This connection between the EEF and DFT provides a different interpretation for the Hartree-exchange-correlation potential $v^{\rm HXC}$ and the Pauli potential $v^{\rm P}$.
  In particular, we showed that $v^{\rm P}$ contains the geometric potential of the KS system, $v^{\rm PG}$, and the energy of the KS environment, $v^{\rm PH}$, whereas $v^{\rm HXC}$ can be viewed as a correction to the external potential due to the different electron-electron interaction in the KS system compared to the interacting system.
  This has to be contrasted with the usual view of the Pauli potential as being the difference of a non-interacting fermionic and bosonic system -- from the EEF perspective, both $v^{\rm HXC}$ the Pauli potential describe the fermionic problem itself, just in a different way than how the interacting many-electron wavefunction describes the problem.
  
  In contrast to $v^{\rm H}$, the physical meaning of the geometric potential $v^{\rm G}$ is less obvious.
  In this work, we explained its connection to the quantum geometric tensor and showed that $v^{\rm G}$ is a metric that represents changes of the state $\phi$ of the electrons in the environment.
  We illustrated that meaning by translating or/and scaling $\phi(x_2;x_1)$ for a one-dimensional model, and we found that $v^{\rm G}(x_1)$ reflects these transformations.
  The faster $\phi(x_2;x_1)$ changes with $x_1$, the higher the amplitude of $v^{\rm G}(x_1)$ becomes in the regions where the change of $\phi$ happens.
  
  Then, we studied the behavior of $v^{\rm G}(x_1)$ for a one-dimensional model of a two-electron homo- and heteronuclear diatomic molecule.
  Whenever there is a rearrangement of the electron in the environment, this is clearly visible as a peak in $v^{\rm G}(x_1)$.
  The model can be understood in a two-state picture where the wavefunction of the environment $\phi$ changes from one state to the other, which allowed us to provide an analytical form of $v^{\rm G}(x_1)$ as well as constraints on the integral of $\sqrt{v^{\rm G}}$ if a state change happens.
  These results can be useful for more general systems, because also those can often be described by a two-state approach, for example, if an electron hops from one KS orbital to another.
  Such situations reflect charge transfer processes and they are common in molecules, hence further investigation on the behavior of $v^{\rm G}$ can help to model its contribution in the EEF and in DFT.
  
  What we largely avoided in our discussion is the role of the vector potential, which is also part of the geometric picture of the EEF.
  Although it may not be needed to describe the ground state of a many-electron system, it will certainly be relevant for rotating molecules, molecules in laser fields and possibly for describing degenerate states in molecules.
  To investigate the vector potential and to illuminate the importance of the geometric potential further, however, it is necessary to look at three-dimensional model systems, which poses an interesting challenge for future work.

  \section*{Acknowledgement}
    
  AS thanks Denis Jelovina (ETH Z\"urich) for helpful discussions.
  This research is supported by an Ambizione grant of the Swiss National Science Foundation (SNF).
  
  \bibliography{lit}{}
  \bibliographystyle{unsrt}
  
  \appendix
  
  \section{Other expressions for the potentials in the EEF}
    \label{sec:eefpot}
  
    Relations \eqref{eq:eef_vt}, \eqref{eq:eef_vv}, and \eqref{eq:eef_vfs} for the potentials $v^{\rm T}$, $v^{\rm V}$, and $v^{\rm G}$, respectively, can be expressed in terms of the many-electron wavefunction $\psi$ and the one-electron density $\rho = |\chi|^2$ by using the relation \eqref{eq:phi} for the conditional wavefunction $\phi$.
    We first note that the geometric potential is also given by 
    \begin{align}
      v^{\rm G}(\vec{r}) 
        &= \frac{1}{2} \braket{\prv \phi(2,\dots,N;\vec{r}) | \hat{P} | \prv \phi(2,\dots,N;\vec{r})}_{2 \dots N},
        \label{eq:eef_vfs2}
    \end{align}
    where 
    \begin{align}
      \hat{P} = 1 - \ket{\phi(2,\dots,N;\vec{r})}\bra{\phi(2,\dots,N;\vec{r})}
    \end{align}
    is a projection operator on the state orthogonal to $\phi$ for a given value of $\vec{r}$.
    Expression \eqref{eq:eef_vfs2} shows the close connection to the Fubini-Study metric \cite{provost1980} and the geometric meaning of $v^{\rm G}$.
    
    Using \eqref{eq:phi}, it is straightforward to show that 
    \begin{align}
      v^{\rm T}(\vec{r})  
        &= \frac{1}{\rho(\vec{r})} 
          \matrixel{\psi(2,\dots,N;\vec{r})}{-\sum\limits_{j=2}^N \frac{\nabla_j^2}{2}}{\psi(2,\dots,N;\vec{r})}_{2 \dots N} \label{eq:vt_psi} \\
      v^{\rm V}(\vec{r})  
        &= \frac{1}{\rho(\vec{r})}
          \matrixel{\psi(2,\dots,N;\vec{r})}{V(1,2,\dots,N)}{\psi(2,\dots,N;\vec{r})}_{2 \dots N} - v^{\rm ext}(\vec{r})
          \label{eq:vv_psi} \\
      v^{\rm G}(\vec{r}) 
        &= \frac{1}{2 \rho(\vec{r})} \braket{\prv \psi(2,\dots,N;\vec{r}) | \hat{P}_{\psi} | \prv \psi(2,\dots,N;\vec{r})}_{2 \dots N},
        \label{eq:vg_psi}
    \end{align}
    with 
    \begin{align}
      \hat{P}_{\psi} = 1 - \frac{1}{\rho(\vec{r})} \ket{\psi(2,\dots,N;\vec{r})}\bra{\psi(2,\dots,N;\vec{r})}.
    \end{align}
    Also $\hat{P}_{\psi}$ is a projector for a given value of $\vec{r}$, i.e., in the subspace of the coordinates $\vec{r}_2, \dots, \vec{r}_N$.
    As only $\psi$ and $\rho$ appear in \eqref{eq:vt_psi}, \eqref{eq:vv_psi}, and \eqref{eq:vg_psi}, but not $\chi$, it is clear that $v^{\rm T}$, $v^{\rm V}$, and $v^{\rm G}$ do not depend on the gauge, i.e., on the choice of the phase of $\chi$.

  \section{The geometric potential and the kinetic energy density}
    \label{sec:ekind}
    
    The difference between the geometric potential $v^{\rm G}$ of the interacting system and the geometric potential $v^{\rm PG}$ of the KS system can, by construction, be related to the different kinetic energy densities.
    For the gauge $\vec{A} \stackrel{!}{=} 0$ and for the case $\chi = \sqrt{\rho} \in \mathbb{R}$, $\phi \in \mathbb{R}$, 
    \begin{align}
      v^{\rm G}(\vec{r})
        &= \frac{1}{2} \braket{\left(\prv \phi(2,\dots,N|\vec{r})\right)^2}_{2 \dots N}
        = \frac{1}{2} \Braket{\left(\prv \frac{\psi(\vec{r},2,\dots,N)}{\chi(\vec{r})}\right)^2}_{2 \dots N} \\
        &= \frac{1}{2} 
            \left(  \frac{\braket{(\prv \psi)^2}_{2 \dots N}}{\chi^2} 
                    - \frac{\braket{\prv (\psi^2)}_{2 \dots N}}{\chi^3} \prv \chi
                    + \frac{(\prv \chi)^2}{\chi^2}
            \right) \\
        &=  \frac{t(\vec{r})}{\chi^2} 
        - \frac{1}{2} \frac{\prv \rho}{\rho} \frac{\prv \chi}{\chi} 
        + \frac{1}{2} \frac{(\prv \chi)^2}{\chi^2},
        \label{eq:vge}
    \end{align}
    where 
    \begin{align}
      t(\vec{r}) = \frac{1}{2}\braket{(\prv \psi)^2}_{2 \dots N} 
    \end{align}
    is the positively-defined one-electron kinetic energy density of the interacting system.
    
    For the KS system a similar relation holds, i.e., 
    \begin{align}
      v^{\rm PG}(\vec{r}) = v^{\rm G}[\varphi^{\rm KS}] 
        &= \frac{1}{2} \sum_{n=1}^N \left( \prv \frac{\vp_n^{\rm KS}(\vec{r})}{\chi(\vec{r})} \right)^2
        = \frac{1}{2} \sum_{n=1}^N 
            \left( \frac{\prv \vp_n^{\rm KS}}{\chi} - \frac{\vp_n^{\rm KS} \prv \chi}{\chi^2}
            \right)^2 \\
        &= \frac{1}{2} \sum_{n=1}^N  \left( \frac{(\prv \vp_n^{\rm KS})^2}{\chi^2} - \frac{\prv \left((\vp_n^{\rm KS})^2\right)}{\chi^3} \prv \chi + \frac{(\vp_n^{\rm KS})^2 }{\chi^4} (\prv \chi)^2 \right) \\
        &= \frac{t^{\rm KS}(\vec{r})}{\chi^2} - \frac{1}{2} \frac{\prv \rho}{\rho} \frac{\prv \chi}{\chi} + \frac{1}{2} \frac{(\prv \chi)^2}{\chi^2} 
        \label{eq:vgpe}
    \end{align}
    where 
    \begin{align}
      t^{\rm KS}(\vec{r}) = \frac{1}{2}\sum_{n=1}^N (\prv \vp_n^{\rm KS})^2
    \end{align}
    is the positively-defined one-electron kinetic energy density of the KS system.
    As the densities in the interacting and the KS systems are the same, the geometric potential $v^{\rm G}$ (Eq.\ \eqref{eq:vge}) and the KS geometric potential $v^{\rm PG}$ (Eq.\ \eqref{eq:vgpe}) differ only by two kinetic energy densities $t$ and $t^{\rm KS}$.

  \section{EEF potentials and the electron correlation}
    \label{sec:ecorr}
    
    Both our model of the homo- and of the heteronuclear diatomic discussed in Sec.\ \ref{sec:dia} are well-described with the Hartree-Fock approximation, i.e., with a system of non-interacting electrons.
    The KS orbitals are thus very similar to the Hartree-Fock orbitals.
    In our experience, this is typical for for one-dimensional models and it is hard to make Hartree-Fock ``fail''.
    One possibility would be to consider the limit of strictly correlated electrons, as was done in \cite{giarrusso2018}.
    Here, we test a different approach:
    We make the electron correlation in our model Hamiltonian more relevant by setting $c_{\rm en}$ to a large value. This results in a broad external potential $v_{\rm en}$, while keeping the electron-electron interaction $v_{\rm ee}$ sharply localized.
    
    \begin{figure}[!htbp]
      \centering
      \includegraphics[width=.5\linewidth]{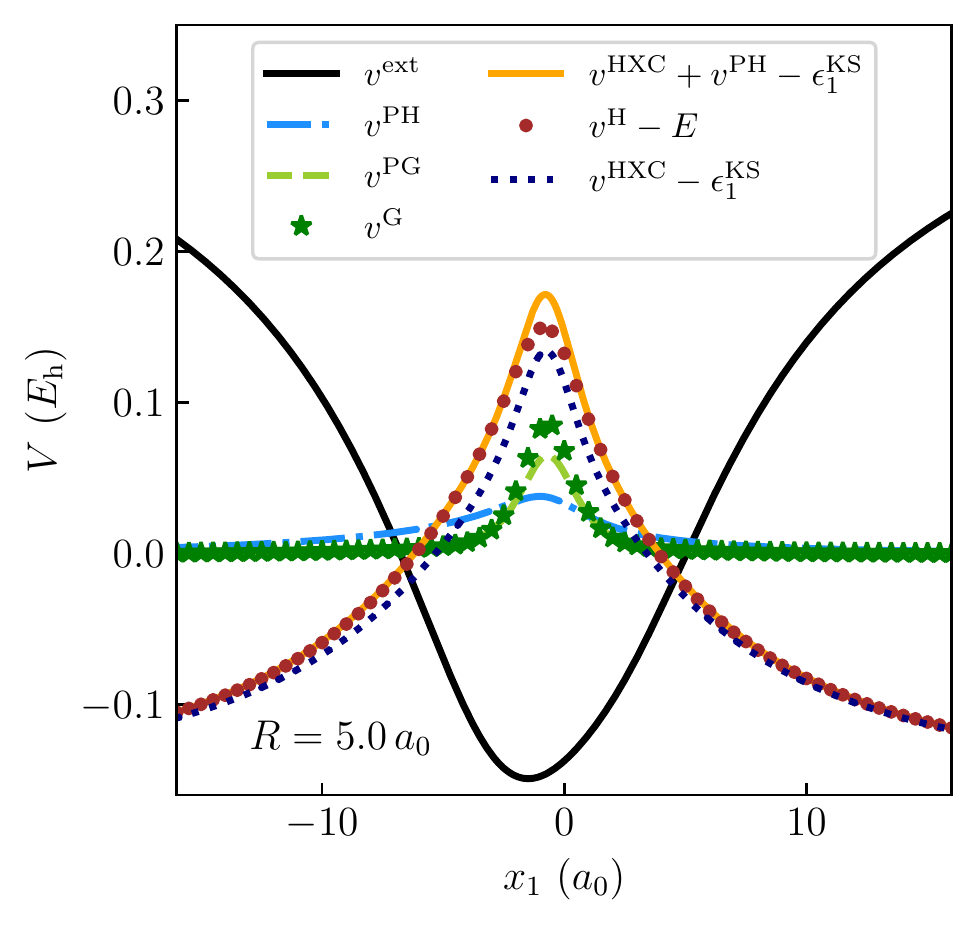}
      \caption{
      Like Fig.\ \ref{fig:pic_geo_asy}d, but for a very delocalized electron-nuclear interaction (parameter \unit[$c_{\rm en} = 25$]{$a_0^2$}).
      }
      \label{fig:pic_non_hf_only_pot}
    \end{figure}
    
    In Fig.\ \ref{fig:pic_non_hf_only_pot}, we show the EEF and DFT potentials for such a large value of $c_{\rm en}$ for the heteronuclear model.
    We find that the geometric potential $v^{\rm G}$ of the interacting wavefunction is higher that the geometric potential $v^{\rm PG}$ of the KS wavefunction, indicating that there is a stronger change of the conditional wavefunction (the state of the electrons in the environment) along $x_1$ in the internuclear region for the interacting system as compared to the KS system.
    As $v$ is the same for the interacting and the KS system, this has to be compensated by $v^{\rm H}$ being smaller than $v^{\rm PH} + v^{\rm HXC}$, i.e., the energy of the environment is lower for the interacting than for the KS system.
    However, we do not think that this test allows for any general conclusions.

\end{document}